\documentclass[journal]{IEEEtran}
\usepackage{amsmath}
\usepackage{amsfonts}
\usepackage{latexsym}
\usepackage{amssymb}
\usepackage{cite}

\usepackage[dvips]{graphicx}
\graphicspath{{./figures/}}
\DeclareGraphicsExtensions{.eps}

\usepackage{array}
\usepackage{theorem}
\usepackage{mathpazo}
\usepackage{times}
\usepackage[makeroom]{cancel}

\usepackage{aliascnt} % allows for a single counter in \autoref'ed theorems
\usepackage[hidelinks,linktocpage]{hyperref}
\usepackage[vlined,boxruled]{algorithm2e}
\SetKwInOut{Input}{input}
\SetKwInOut{Output}{output}
\SetKw{KwDownto}{down to}
\SetKw{KwGoTo}{go to}
\IncMargin{1.2em}
\LinesNumbered

\SetAlFnt{\scriptsize}
\SetAlCapFnt{\scriptsize}
\SetAlCapNameFnt{\scriptsize}

\makeatletter
\renewcommand{\algocf@caption@boxruled}{%
  \hrule
  \hbox to \hsize{%
    \vrule\hskip-0.4pt
    \vbox{   
       \vskip\interspacetitleboxruled%
       \unhbox\algocf@capbox\hfill
       \vskip\interspacetitleboxruled
       }%
     \hskip-0.4pt\vrule%
   }\nointerlineskip%
}%
\makeatother
\usepackage{psfrag}
\usepackage{tikz}
\usetikzlibrary{positioning,decorations.pathmorphing,shapes}
\newcommand{\mynewtheorem}[2]{
  \newaliascnt{#1}{dummy}
  \newtheorem{#1}[#1]{#2}
  \aliascntresetthe{#1}
  \expandafter\def\csname #1autorefname\endcsname{#2}
}

\theoremstyle{plain}
\theorembodyfont{\normalfont}

  \mynewtheorem{theorem}{Theorem$\!$}
  \mynewtheorem{lemma}{Lemma$\!$}
  \mynewtheorem{corollary}{Corollary$\!$}
  \mynewtheorem{xmpl}{Example$\!$}
  \newenvironment{example}{\begin{xmpl}\hspace*{-1ex}}{\hfill$\Box$\end{xmpl}}
  
\newtheorem{cnstr}{Construction$\!$}

\newenvironment{construction}{\begin{cnstr}}{\hfill$\Box$\end{cnstr}}

\newtheorem{step}{Step$\!$}

\newcounter{enumrom}
\renewcommand{\theenumrom}{(\roman{enumrom})}

\makeatletter
\renewcommand{\@endtheorem}{\endtrivlist}
\makeatother

\makeatletter
\renewcommand{\thefigure}{{\@arabic\c@figure}}
\renewcommand{\fnum@figure}{{\bf Figure\,\thefigure}}
\makeatother

\newcommand{\cK}{\mathcal{K}}

\newcommand{\cR}{\mathcal{R}}

\newcommand{\mathset}[1]{\left\{#1\right\}}
\newcommand{\mathsetp}[2]{\mathset{#1 ~|~ #2}}
\newcommand{\abs}[1]{\left|#1\right|}

\newcommand{\parenv}[1]{\left( #1 \right)}
\newcommand{\sparenv}[1]{\left[ #1 \right]}

\newcommand{\ceilenv}[1]{\left\lceil #1 \right\rceil}
\newcommand{\floorenv}[1]{\left\lfloor #1 \right\rfloor}
\newcommand{\tends}[1]{\operatorname*{\longrightarrow}\limits_{#1}}
\newcommand{\vin}{\rotatebox[origin=c]{-90}{$\in$}}
\newcommand{\arrowback}{\rotatebox[origin=c]{-90}{$\curvearrowright$}}
\newcommand\myatop[2]{\genfrac{}{}{0pt}{}{#1}{#2}}
\renewcommand{\Bbb}{\mathbb}

\newcommand{\F}{{\Bbb F}}

\newcommand{\N}{{\Bbb N}}
\newcommand{\R}{{\Bbb R}}
\newcommand{\Z}{{\Bbb Z}}
\newcommand{\quo}[1]{``#1''}
\newcommand{\pttt}{\quo{push-to-the-top} }

\newcommand{\Gaux}{G^{\operatorname{aux}}_{\uparrow}}
\newcommand{\Caux}{C^{\operatorname{aux}}}

\newcommand{\shat}{\hat{\sigma}}
\newcommand{\phat}{\hat{\pi}}
\DeclareMathOperator*{\argmin}{argmin}
\DeclareMathOperator{\id}{Id}

\renewcommand{\leq}{\leqslant}

\renewcommand{\geq}{\geqslant}

\begin{document}
%%%%%%%%%%%%%%%%%%%%%%%%%%%%%%%%%%%%%%%%%%%%%%%%%%%%%%%
%%%%%%%%%%%%%%%%%%%%%%%%%%%%%%%%%%%%%%%%%%%%%%%%%%%%%%%
\title{Limited-Magnitude Error-Correcting Gray Codes for Rank Modulation}
%%%%%%%%%%%%%%%%%%%%%%%%%%%%%%%%%%%%%%%%%%%%%%%%%%%%%%%
%%%%%%%%%%%%%%%%%%%%%%%%%%%%%%%%%%%%%%%%%%%%%%%%%%%%%%%
\author{Yonatan~Yehezkeally,~\IEEEmembership{Student~Member,~IEEE,}
        and~Moshe~Schwartz,~\IEEEmembership{Senior~Member,~IEEE}% <-this % stops a space
        \thanks{This work was supported in part by ISF grant no.~130/14. This paper will be presented in part at the 2016 IEEE International Symposium on Information Theory.}%
        \thanks{The authors are with the Department of Electrical and Computer Engineering, Ben-Gurion University of the Negev, Beer Sheva 8410501, Israel
  (e-mail: yonatany@post.bgu.ac.il; schwartz@ee.bgu.ac.il).}
}
%%%%%%%%%%%%%%%%%%%%%%%%%%%%%%%%%%%%%%%%%%%%%%%%%%%%%%%
\maketitle
\begin{abstract}
  We construct Gray codes over permutations for the rank-modulation
  scheme, which are also capable of correcting errors under the
  infinity-metric. These errors model limited-magnitude or spike
  errors, for which only single-error-detecting Gray codes are
  currently known. Surprisingly, the error-correcting codes we
  construct achieve a better asymptotic rate than that of
  presently known constructions not having the Gray property, and
  exceed the Gilbert-Varshamov bound. Additionally, we present
  efficient ranking and unranking procedures, as well as a decoding
  procedure that runs in linear time. Finally, we also apply our 
  methods to solve an outstanding issue with error-detecting 
  rank-modulation Gray codes (snake-in-the-box codes) under a different 
  metric, the Kendall $\tau$-metric, in the group of permutations over 
  an even number of elements $S_{2n}$, where we provide asymptotically 
  optimal codes.
\end{abstract}
%%%%%%%%%%%%%%%%%%%%%%%%%%%%%%%%%%%%%%%%%%%%%%%%%%%%%%%
%%%%%%%%%%%%%%%%%%%%%%%%%%%%%%%%%%%%%%%%%%%%%%%%%%%%%%%
%%%%%%%%%%%%%%%%%%%%%%%%%%%%%%%%%%%%%%%%%%%%%%%%%%%%%%%%%%%%%%%%%%%%%%
%%%%%%%%%%%%%%%%%%%%%%%%%%%%%%%%%%%%%%%%%%%%%%%%%%%%%%%%%%%%%%%%%%%%%%

\section{Introduction}

\IEEEPARstart{R}{ank} modulation is a method for storing information
in non-volatile memories \cite{JiaMatSchBru09}, which has been
researched in recent years. It calls for storing information in
relative values stored in a group of cells rather than the absolute
values of single cells. More precisely, it stores information in the
permutation suggested by sorting a group of cells by their relative
values, e.g., charge levels in flash memory cells. It allows for
increased robustness against certain noise mechanisms (e.g., charge
leakage in flash memory cells), as well as alleviating some inherent
challenges in flash memories (e.g., programming/erasure-asymmetry and
programming-overshoot). Permutation codes have also previously seen
usages in source-encoding \cite{Sle65,BerJelWol72,Bla74,BlaCohDez79}
and signal detection \cite{ChaKur69}, as well as other fields
\cite{ChaRee70,CohDez77,DezFra77}, and more recently been used in
power-line communications \cite{VinHaeWad00}.

Several error models have been studied for rank modulation, including
the Kendall $\tau$-metric
\cite{JiaSchBru10,BarMaz10,MazBarZem13,ZhoSchJiaBru15}, the
$\ell_\infty$-metric \cite{TamSch10,KloLinTsaTze10,ShiTsa10,TamSch12}
and other examples \cite{DinFuKloWei02,FuKlo04}. In this paper we
focus on the $\ell_\infty$-metric, which models limited-magnitude or
spike noise, i.e., we assume that the rank of any given cell--its
position when sorting the group of cells--could not have changed by
more than a given amount. \cite{TamSch10,KloLinTsaTze10} have
presented constructions for error-correcting codes under this metric,
as well as explored some non-constructive lower- and upper-bounds on
the parameters of existing codes. \cite{TamSch12} has since employed
methods of relabeling to optimize the minimal distance of known
constructions.

Gray codes were first discussed over the space of binary vectors, 
where each pair of consecutive vectors differed by a single bit-flip
\cite{Gra53}; the concept has since been generalized to include codes
over arbitrary alphabets, requiring only that codewords could be
ordered in a sequence, where each codeword is derived from the
previous by one of a predefined set of functions. Other suggested
usages of Gray codes, surveyed in \cite{Sav97}, include
permanent-computation \cite{NijWil87}, circuit-testing
\cite{RobCoh81}, image-processing \cite{AmaSunDha90}, hashing
\cite{Fal88}, coding \cite{Etz92,SchEtz99,JiaMatSchBru09} and data
storing/extraction \cite{ChaCheChe92}. In particular, in the context
of rank modulation, the use of Gray codes has been shown to reduce
write-time by eliminating the risk of programming-overshoot and allow
integration with other multilevel-cells coding schemes
\cite{JiaMatSchBru09,EngLanSchBru11b,EngLanSchBru13}.

Gray codes with error-correction capabilities have sometimes been
referred to as spread-$d$ circuit codes (see \cite{HooRecSawWon13} and
references therein). Specifically, in the context of rank modulation,
such codes were so far only studied for the case of
single-error-detection, where they were dubbed snake-in-the-box codes
(or, more appropriately, coil-in-the-box codes, when they are
cyclic). \cite{YehSch12b} studied these codes under both the Kendall
$\tau$-metric and the $\ell_\infty$-metric, and more recent papers
\cite{HorEtz14,ZhaGe16,Hol16} have categorized and constructed 
optimally sized coil-in-the-box codes under the former metric for odd 
orders, although the case of even orders proved more challenging (see 
\cite{ZhaGe17} in addition to the aforementioned papers).

In this work we focus on the $\ell_\infty$-metric and present a
construction of error-correcting Gray codes capable of correcting an
arbitrary number of limited-magnitude errors. The allowed transitions
between codewords are the \quo{push-to-the-top} operations, used in most
previous works
\cite{JiaMatSchBru09,YehSch12b,HorEtz14,ZhaGe16,Hol16,EngLanSchBru11b,EngLanSchBru13}.
The resulting codes will be shown to be larger than known constructions 
in the case of fixed minimal distance, as well as achieve better
asymptotic rates than known codes in the case of $d=\Theta(n)$.

We will also briefly examine error-detecting codes under the Kendall 
$\tau$-metric for even orders, since methods developed for the 
application of our main construction can readily be adapted to that 
purpose. We provide an asymptotically optimal construction which nearly 
completes the categorization of available codes for that scheme.

The paper is organized as follows. In \autoref{sec:prel} we present
notations and definitions. In \autoref{sec:auxcodes} we study a new 
kind of auxiliary codes which are required for our construction, before 
presenting it in \autoref{sec:const} and discussing its performance in 
comparison with known constructions and bounds. We devise a decoding 
algorithm for the generated codes in \autoref{sec:decodalg}, and discuss 
ranking and unranking procedures in \autoref{sec:ranking}. We briefly 
present an adaptation of the developed auxiliary codes to error-detecting 
codes under the Kendall $\tau$-metric in \autoref{sec:app}. Finally, we 
conclude in \autoref{sec:conclusion} by reviewing our results and 
suggesting problems for future study.

\section{Preliminaries}\label{sec:prel}
For $n\in\N$, we let $S_n$ be the \emph{symmetric group}, the set 
of all permutations on $[n] = \mathset{1,2,\ldots,n}$ (i.e., 
bijections
$\sigma:[n] \operatorname*{\longrightarrow}\limits^{1-1}_{\text{onto}} [n]$), 
with composition as group action:
\[\sigma\tau(k) = \parenv{\sigma\circ\tau}(k) = \sigma(\tau(k)).\]
Throughout the paper we shall denote the identity permutation 
$\id\in S_n$ defined for all $k\in[n]$: $\id(k) = k$.

We use the \emph{cycle notation} for permutations, i.e., for distinct
$\mathset{a_j}_{j=1}^k\subseteq [n]$ we let
$\sigma=(a_1,a_2,\ldots,a_k)$ be the permutation such that
$\sigma(a_j) = a_{\parenv{j\bmod k}+1}$ and $\sigma(b)=b$ for all
$b\in [n]\setminus\mathset{a_j}_{j=1}^k$. Trivially, every permutation
can be represented as a composition of disjoint cycles.  It is also
well known that every permutation can be represented as a composition
of \emph{transpositions}, cycles of length $2$, and that the parity of
the number of transpositions in that representation is unique
(although the representation itself is not). We therefore have
\emph{even} and \emph{odd} permutations, and the set of even
permutations forms a subgroup $A_n\leq S_n$ named the
\emph{alternating group}. We will say that $C\subseteq S_n$ is
\emph{parity-preserving} if every two elements $\sigma,\tau\in C$ have
the same parity, that is, $\operatorname*{sign}\sigma = 
\operatorname*{sign}\tau$.

We also use the \emph{vector notation} for permutations,
\[\sigma = \sparenv{\sigma(1),\ \sigma(2),\ \ldots\ , \sigma(n)}.\]
This allows us to more easily notate, for $1\leq i<j\leq n$, the 
\emph{\quo{push-to-the-$i$th-index}} transition 
$t_{i\uparrow j}:S_n\to S_n$ by
\begin{align*}
t_{i\uparrow j}\parenv{\sparenv{a_1,a_2,\ldots,a_{i-1},a_i,a_{i+1},
\ldots,a_{j-1},a_j,a_{j+1},\ldots,a_n}}& \\
= \sparenv{a_1,a_2,\ldots,a_{i-1},a_j,a_i,a_{i+1},\ldots,
a_{j-1},a_{j+1},\ldots,a_n}.&
\end{align*}

We follow \cite{JiaMatSchBru09,YehSch12b,HorEtz14,ZhaGe16} (among others) 
in dubbing \quo{push-to-the-$1$st-index} transitions as \emph{\pttt} 
transitions, and we denote $t_{\uparrow j}=t_{1\uparrow j}$.
Finally, we define the \emph{\quo{push-to-the-bottom}} 
transition on the $j$th index, $t_{\downarrow j}:S_n\to S_n$,
\begin{align*}
t_{\downarrow j}\parenv{\sparenv{a_1,a_2,\ldots,a_{j-1},a_j,a_{j+1},\ldots,a_n}}& \\
= \sparenv{a_1,a_2,\ldots,a_{j-1},a_{j+1},\ldots,a_n,a_j}.&
\end{align*}

Given any set $S$, and a collection of transitions
\[T \subseteq \mathsetp{f}{f:S\to S},\]
we define a \emph{$T$-Gray code} over $S$ to be a sequence ${C = 
\parenv{c_r}_{r=1}^M\subseteq S}$ such that for all $1\leq r<r'\leq
M$ we have $c_r\neq c_{r'}$ and such that for all $1\leq r<M$ there
exists $t_r\in T$ satisfying $c_{r+1}=t_r(c_r)$ (we say that a
sequence $C$ is contained in $S$, by abuse of notation, if $c_r\in S$
for all $r$. That is, we may refer to a Gray Code as an unordered
set--or simply a code--when desired for simplicity). We call
$M=\abs{C}$ the \emph{size} of the code, and $t_1,t_2,\ldots, t_{M-1}$
the transition sequence \emph{generating} $C$. If there exists $t\in
T$ such that $c_1=t(c_M)$ we say that $C$ is \emph{cyclic}, and
include $t_M=t$ in its generating transition sequence. If $C=S$, we
say that $C$ is a \emph{complete} code.

\begin{example}
In the classic example of a Gray code we have, e.g., $S=\F_2{}^3$, 
with $T$ consisting of the group action of 
$\mathset{001, 010, 100}\subseteq S$ on $S$, defined 
\[v(u) = u+v.\]
We then have the complete cyclic Gray code given by
\[
000
\substack{001\\ \longrightarrow\\ \ }
001
\substack{010\\ \longrightarrow\\ \ }
011
\substack{001\\ \longrightarrow\\ \ }
010
\substack{100\\ \longrightarrow\\ \ }
110
\substack{001\\ \longrightarrow\\ \ }
111
\substack{010\\ \longrightarrow\\ \ }
101
\substack{001\\ \longrightarrow\\ \ }
100
\substack{100\\ \longrightarrow\\ \ }
000.
\]
\end{example}

In this paper, we fix $S=S_n$. We say that $C=
\parenv{c_1,c_2,\ldots,c_M}\subseteq S_n$ is a
$G_{i\uparrow}(n,M)$ if it is a cyclic Gray code with transition set
$T = \mathsetp{t_{i\uparrow j}}{i<j\leq n}$. When $i=1$ we refer to
$C$ as a \pttt code and denote it $G_{\uparrow}(n,M)$, and we likewise
denote \quo{push-to-the-bottom} codes $G_{\downarrow}(n,M)$.

\begin{example}
It has been remarked in \cite{JiaMatSchBru09} that
\[
\begin{bmatrix}
1\\2\\3
\end{bmatrix}
\operatorname*{\longrightarrow}^{t_{\uparrow 2}}
\begin{bmatrix}
2\\1\\3
\end{bmatrix}
\operatorname*{\longrightarrow}^{t_{\uparrow 3}}
\begin{bmatrix}
3\\2\\1
\end{bmatrix}
\operatorname*{\longrightarrow}^{t_{\uparrow 3}}
\begin{bmatrix}
1\\3\\2
\end{bmatrix}
\operatorname*{\longrightarrow}^{t_{\uparrow 2}}
\begin{bmatrix}
3\\1\\2
\end{bmatrix}
\operatorname*{\longrightarrow}^{t_{\uparrow 3}}
\begin{bmatrix}
2\\3\\1
\end{bmatrix}
\operatorname*{\arrowback}^{t_{\uparrow 3}}
\]
is a $G_\uparrow(3,6)$, i.e., a complete cyclic \pttt Gray code 
over $S_3$.
\end{example}

It is worthwhile to note that when $S$ is a group, and $T$ consists of
the group action of some subset on $S$, and $C$ is a (complete- and/or
cyclic-) Gray code generated by $t_1,t_2,\ldots, t_{M-1}$ ($,t_M$),
then for all $\sigma\in S$ we observe that
$\parenv{\sigma,t_1(\sigma), t_2(t_1(\sigma)),\ldots}$ is also a
(complete- and/or cyclic- respectively) Gray code.  In other words,
the code is shift invariant. In these cases we might refer to the
transition sequence generating the code as the code itself, when
desirable for simplicity. It is also of interest to observe that
$t_{i\uparrow j}(\sigma) = \sigma\circ (j,j-1,\ldots,i)$, i.e.,
\quo{push-to-the-$i$th-index} transitions are group actions.

When $S$ is equipped with a metric $d:S\times S\to \R_+$, and
$C\subseteq S$ has the property that for all $\sigma,\tau\in C$ either
$\sigma=\tau$ or $d(\sigma,\tau)\geq d$, for some constant $d>0$, then
$C$ (when considered as an unordered set) is commonly referred to as
an \emph{error-correcting code} with \emph{minimal distance} $d$. If
$d(\cdot,\cdot)$ models an error mechanism, such that a single error
corresponds to distance $1$, and $2p+q < d$, it is well known that $C$
can then correct $p$ errors, and also detect $q$ additional errors.

Error-correcting Gray codes have sometimes been referred to as
\emph{spread-$d$ circuit codes} (see \cite{HooRecSawWon13} and
references therein), where they were traditionally defined by
requiring that for all $c_r,c_{r'}\in C$, $(r-r'\bmod \abs{C})\geq d$
implies $d(c_r,c_{r'})\geq d$. In that way, e.g., spread-$1$
circuit codes are traditional Gray codes. This eased requirement was
made necessary since, working with the Hamming distance $d_H$ in the
$n$-cube, one cannot have codewords at distance less than $d$ in the
code sequence attain a distance of at least $d$. We shall depart from
it here to deal with Gray codes which are classic error-correcting
codes, but the codes presented in this paper are nevertheless also, in
particular, spread-$d$ circuit-codes.

We shall focus on the $\ell_\infty$-metric defined on $S_n$ by 
\[d_\infty(\sigma,\tau) = \max_{j\in [n]}\abs{\sigma(j)-\tau(j)}.\]
That is, it is the metric induced on $S_n$ by the embedding into 
$\Z^n$ (and, indeed, $\R^n$) implied by the vector notation, and 
the $\ell_\infty$-metric in these spaces.
\cite{TamSch10} studied error-correcting codes in $S_n$ with 
$d_\infty$, which it dubbed \emph{limited-magnitude rank-modulation} 
codes, and denoted a code $C$ with minimal distance $d$ as an 
$(n,\abs{C},d)$-LMRM code. In our case, if a $G_{i\uparrow}(n,M)$ 
is also an $(n,M,d)$-LMRM code, we shall denote it a 
$G_{i\uparrow}(n,M,d)$ (likewise for $G_\uparrow$ and $G_\downarrow$).

Finally, we organize our notation of codes in \autoref{codenotate}.

\begin{table}[t]
\caption{Code notations for $C\subseteq S_n$.}
\label{codenotate}
\vspace{-2.0em}
\begin{center}
%\bgroup
\renewcommand{\arraystretch}{1.5}
\begin{tabular}{m{0.25\linewidth}m{0.6\linewidth}}
Notation & Definition\\ \hline\hline
$G_{i\uparrow}(n,M)$ & $C=\parenv{c_r}_{r=1}^M\subseteq S_n$ such that 
for all $r$:\newline ${c_{(r\bmod M)+1}=t_{i\uparrow j_r}(c_r)}$.\\ \hline
$G_{\uparrow}(n,M)$ & $C$ is a $G_{1\uparrow}(n,M)$.\\ \hline
$G_{\downarrow}(n,M)$ & $C=\parenv{c_r}_{r=1}^M\subseteq S_n$ such that 
for all $r$:\newline $c_{(r\bmod M)+1}=t_{\downarrow j_r}(c_r)$.\\ \hline
$(n,M,d)$-LMRM & $C\subseteq S_n$, $\abs{C}=M$ and 
for all $c_1\neq c_2\in C$:\newline $d_\infty(c_1,c_2)\geq d$.\\ \hline
${G_{i\uparrow}(n,M,d)}$ & $C$ is a $G_{i\uparrow}(n,M)$ and an $(n,M,d)$-LMRM.\\ \hline
${G_{\uparrow}(n,M,d)}$ & $C$ is a $G_{1\uparrow}(n,M,d)$.\\ \hline
$\Gaux(k,M)$ & $C$ is a $G_{\uparrow}(k,M)$, 
and for all ${q\in[k-1]}$: ${\sigma\in C}\implies (q,k)\sigma\not\in C$.\newline
(See \autoref{sec:auxcodes}.)
\end{tabular}
%\egroup
\end{center}
\vspace{-2.0em}
\end{table}

\section{Auxiliary construction}\label{sec:auxcodes}

Before we present the main construction of our paper, we first describe 
in this section a construction for auxiliary codes which will be a 
component of the main construction.

We define auxiliary codes in $S_k$ in the following way: we say 
that $C$ is $\Gaux(k,M)$ if it is 
$G_{\uparrow}(k,M)$ and for all $q\in[k-1]$ it holds that 
\[\sigma\in C \implies (q,k)\circ\sigma\not\in C.\]
In our main construction, we will use a $\Gaux(k,M)$ code 
$\parenv{c_r}_{r=1}^M$, which we also require to satisfy $c_1 = \id$ 
and $c_2 = t_{\uparrow k}\id$. We hence study the existence of such 
codes.

Firstly, note that the only existing $\Gaux(2,M)$ codes 
are the singletons $\mathset{\id},\mathset{(1,2)}$. However, for 
$k\geq 3$ there do exist $\Gaux(k,M)$ codes with $M\geq 3$, 
as one such example is 
\[\mathset{\id, t_{\uparrow 3}\id, t_{\uparrow 3}{}^2\id}.\]
We also note the following:
\begin{lemma}
If $C\subseteq S_k$ is $\Gaux(k,M)$, then $M\leq\frac{\abs{S_k}}{2}$.
\end{lemma}
\begin{IEEEproof}
Take $q\in[k-1]$, and observe that $\sigma\mapsto (q,k)\sigma$ 
is an $S_k$-automorphism, under which $C$ and its image are disjoint. 
Hence $2M\leq\abs{S_k}$.
\end{IEEEproof}

This motivates us to examine another family of codes, namely, 
parity-preserving codes, due to the following observations.
\begin{lemma}
If $C\subseteq S_k$ is parity-preserving then 
$\abs{C}\leq\frac{\abs{S_k}}{2}$.
\end{lemma}
\begin{IEEEproof}
Either $C\subseteq A_k$ or $C\subseteq S_k\setminus A_k$.
\end{IEEEproof}
\begin{lemma}
  \label{lem:par-pres-aux}
  If $C\subseteq S_k$ is a parity-preserving $G_\uparrow(k,M)$, then 
  $C$ is $\Gaux(k,M)$.
\end{lemma}
\begin{IEEEproof}
Observe that $\operatorname*{sign}\sigma \neq \operatorname*{sign}
(q,k)\sigma$ for all ${q\in[k-1]}$, thus both cannot belong to $C$.
\end{IEEEproof}

Parity-preserving $\Gaux(2m+1,M)$ codes are known to exist,
achieving the aforementioned bound.
\begin{lemma}\label{holsnakes}\cite{Hol16}
For all $m\neq 2$, there exist parity-preserving 
$G_\uparrow(2m+1,\frac{(2m+1)!}{2})$ codes. The largest parity-preserving 
$G_\uparrow(5, M)$ codes have $M=57$.
\end{lemma}

Although not declared, it is shown in \cite{Hol16} that such codes can 
be assumed to have $t_{\uparrow 2m+1}$ as the first transition in
their generating transition sequence, and furthermore, that they also 
employ at least one $t_{\uparrow 2m-1}$ transition.

In comparison, as noted in \cite{YehSch12b}, a parity-preserving
$G_\uparrow(2m,M)$ must satisfy $M\leq\frac{\abs{S_{2m}}}{2m}$, as it
must never employ a $t_{\uparrow 2m}$ transition. We therefore examine
more general $\Gaux$ codes, which are not
parity-preserving. We begin by noting the following lemma.

\begin{lemma}\label{jsnakes}\cite[Thm.~4,7]{JiaMatSchBru09}
For all $n\geq 1$ there exist $G_\uparrow(n,n!)$ codes, that is, 
complete and cyclic \pttt Gray codes over the symmetric 
group $S_n$.
\end{lemma}

Relying on these codes, we construct auxiliary codes in the following
theorem. Similarly constructed codes already appeared in a different 
context as components in a construction in \cite{JiaMatSchBru09}.

\begin{theorem}\label{flipconst}
For all $m\geq 2$ there exists a
$\Gaux(2m,\frac{\abs{S_{2m}}}{2m-1})$, starting at $\id$ and
with a generating sequence starting with $t_{\uparrow 2m}$.
\end{theorem}
\begin{IEEEproof}
Take a $G_\uparrow(2m-2, (2m-2)!)$ code $C^\prime$, provided by
\autoref{jsnakes}. We follow the concept of
\cite[Thm.~7]{JiaMatSchBru09} in extending $C^\prime$ to
$S_{2m}$. Let us define
\[\sigma_0 = t_{\uparrow 2m}\id = \sparenv{2m,1,\ldots,2m-1}.\]
If we take $t_{\uparrow i_1}, t_{\uparrow i_2},\ldots, 
t_{\uparrow i_{(2m-2)!}}$ to be the
transition sequence generating $C^\prime$, then the transition
sequence 
\[t_{\downarrow 2m+1-i_1}, t_{\downarrow 2m+1-i_2}, 
\ldots, t_{\downarrow 2m+1-i_{(2k-2)!}}\]
of
\quo{push-to-the-bottom} operations, applied in succession to
$\sigma_0$, generates $C^{\prime\prime}\subseteq S_{2m}$, a
$G_\downarrow(2m,(2m-2)!)$, all of whose elements' vector notations
begin with $\sparenv{2m,1}$.

We now note that $t_{\downarrow 2m+1-j} = t_{\uparrow 2m}{}^{2m-1}
t_{\uparrow j}$. Thus, by replacing each $t_{\downarrow 
2m+1-j}$ with $t_{\uparrow j}$ followed by a sequence of $2m-1$
occurrences of $t_{\uparrow 2m}$, we get $C\subseteq S_{2m}$, a
$G_\uparrow(2m,(2m-2)!2m)$, where every block of $2m$ elements is
comprised of cyclic shifts of some $\sigma\in C^{\prime\prime}$.

The code $C$ is known to be a Gray code \cite[Thm.~7]{JiaMatSchBru09}.
Moreover, if $\sigma\in C$ satisfies $\tau=(q,2m)\sigma\in C$, note
that both have a vector notation with $1$ immediately
(cyclically) following $2m$, but since $\tau=(q,2m)\sigma$ its vector
notation has $1$ following $q$. It follows (by abuse of notation) that
$q=2m$.

Finally, note that $C$ is generated by a transition sequence ending 
with $2m-1$ instances of $t_{\uparrow 2m}$, so it includes $\id$ followed 
by a $t_{\uparrow 2m}$ transition. A cyclic shift of $C$ therefore 
satisfies the theorem.
\end{IEEEproof}

\begin{example}\label{exm:aux}
To construct a $\Gaux(4,8)$ we utilize the complete $G_\uparrow(2,2)$ 
code $\mathset{\id, t_{\uparrow 2}\id}$, generated by $t_{\uparrow 2}, 
t_{\uparrow 2}$, to arrive by the $G_\downarrow(4,2)$ code
\[C^{\prime\prime} = \mathset{\sparenv{4,1,2,3}, \sparenv{4,1,3,2}},\]
which is generated by $t_{\downarrow 3}, t_{\downarrow 3}$. We recall 
that $t_{\downarrow 3} = t_{\uparrow 4}{}^3\circ t_{\uparrow 3}$, 
allowing us to expand $C^{\prime\prime}$ in the following manner:
\begin{footnotesize}
\[
\operatorname*{\begin{bmatrix}
4\\ 1\\ 2\\ 3
\end{bmatrix}}_{\myatop{\vin}{C^{\prime\prime}}}
\operatorname*{\to}^{t_{\uparrow 3}}
\begin{bmatrix}
2\\ 4\\ 1\\ 3
\end{bmatrix}
\operatorname*{\to}^{t_{\uparrow 4}}
\begin{bmatrix}
3\\ 2\\ 4\\ 1
\end{bmatrix}
\operatorname*{\to}^{t_{\uparrow 4}}
\begin{bmatrix}
1\\ 3\\ 2\\ 4
\end{bmatrix}
\operatorname*{\to}^{t_{\uparrow 4}}
\operatorname*{\begin{bmatrix}
4\\ 1\\ 3\\ 2
\end{bmatrix}}_{\myatop{\vin}{C^{\prime\prime}}}
\operatorname*{\to}^{t_{\uparrow 3}}
\begin{bmatrix}
3\\ 4\\ 1\\ 2
\end{bmatrix}
\operatorname*{\to}^{t_{\uparrow 4}}
\begin{bmatrix}
2\\ 3\\ 4\\ 1
\end{bmatrix}
\operatorname*{\to}^{t_{\uparrow 4}}
\begin{bmatrix}
1\\ 2\\ 3\\ 4
\end{bmatrix}
\operatorname*{\arrowback}^{t_{\uparrow 4}}
\]
\end{footnotesize}
Finally, we observe that shifting the resulting code so it begins with
$\id$ is satisfactory.
\end{example}

We remark that, while \autoref{flipconst} does not produce auxiliary
codes much larger than the parity-preserving code of size
$\frac{\abs{S_{2m}}}{2m}$, it does at least allow us to permute the
last element.

Next, we present another construction which yields larger codes, for 
even $k\geq 6$ (but not $k=4$). From now on, we fix $m\geq 2$. We also 
define ${\varphi:S_{2m+2}\to S_{2m+2}}$ by
\[\varphi = t_{\uparrow 2m+2}{}^2\circ t_{\uparrow 2m-1}{}^{-1}.\]
We note that 
\[\varphi(\pi) = \pi\circ (1,2m+1)(2m+2,2m,2m-1,\ldots,2),\]
Hence, informally, in $\pi$'s vector notation, $\varphi$ transposes the 
elements in indices $1,2m+1$, and cyclically shifts all other elements 
once to the top. We can also observe that $\varphi^{2m}=\id$.

We conveniently define, for all $r\geq 0$, the permutations
\begin{align*}
\phat_r &= \varphi^r(\id) \\
&= (1,2m+1)^r(2m+2,2m,2m-1,\ldots,2)^r \in S_{2m+2},
\end{align*}
In particular, we note that when $r\equiv r^\prime\pmod{2m}$, and only 
then, we have $\phat_r = \phat_{r^\prime}$.

\begin{lemma}
  \label{aux:cycle}
  For all $r\geq 0$ a parity-preserving 
  $G_\uparrow(2m+2,M_{2m+2})$ code $P_r$ exists 
  which begins with $\phat_r$ and ends with $t_{\uparrow 2m-1}{}^{-1} 
  \phat_r$, where 
  \[M_{2m+2} = \begin{cases}
  57 & m=2,\\
  \frac{(2m+1)!}{2} & m>2.
  \end{cases}\]
\end{lemma}
\begin{IEEEproof}
The claim follows trivially from \autoref{holsnakes}, if we shift the 
generating transition sequence such that it ends with 
$t_{\uparrow 2m-1}$ and apply it to $\phat_r$, due to 
\autoref{lem:par-pres-aux}.
\end{IEEEproof}

We note in particular that for all $r$, $\phat_r$ is even, and thus
$P_r\subseteq A_{2m+2}$. Moreover, since the parity-preserving code $P_r$ 
does not employ $t_{\uparrow 2m+2}$, for all $\pi\in P_r$ it holds that 
\begin{align*}
\pi(2m&+2) = \phat_r(2m+2) \\
&= \begin{cases}
2m+2 & r\equiv 0\pmod{2m},\\
2m+1 - \parenv{r\bmod 2m} & r\not\equiv 0\pmod{2m}.
\end{cases}
\end{align*}
Thus, when considered as sets,
\[ P_r \cap P_{r'} = \emptyset,\]
for all $0\leq r < r' < 2m$.

We shall construct a $\Gaux(2m+2,M)$ code by stitching together
$P_1,P_2,\ldots, P_{2m-1}$. We will need to amend $P_0$ before 
incorporating it into our code, for reasons we shall discuss below. 
First, we describe the stitching method in the following lemma.

\begin{lemma}
  \label{aux:stitch}
  For all $r\geq 0$ (including, in particular, $r = {2m-1}$), we may 
  concatenate $P_r, P_{r+1}$ into a (non-cyclic) 
  \quo{push-to-the-top} code by applying the transitions 
  $t_{\uparrow 2m+2}, t_{\uparrow 2m+2}$ to the last permutation of 
  $P_r$, which is $t_{\uparrow 2m-1}{}^{-1}\phat_r$. 
  Additionally, the only odd permutation in the resulting code is 
  \[\beta_{r+1} = t_{\uparrow 2m+2}{}^{-1}(\phat_{r+1}).\]
  We shall refer to it as the \emph{$(r+1)$-bridge}.
\end{lemma}
\begin{IEEEproof}
The claim follows trivially from the definition
\[\phat_{r+1} = \varphi(\phat_r) = t_{\uparrow 2m+2}\circ
\parenv{t_{\uparrow 2m+2}\circ t_{\uparrow 2m-1}{}^{-1}(\phat_r)},\]
since $P_r,P_{r+1}$ are parity-preserving, and $t_{\uparrow 2m+2}$ 
flips parity.
\end{IEEEproof}

\autoref{aux:stitch} can be used iteratively to concatenate
$P_1,P_2,\ldots, P_{2m-1}$, with a single odd permutation--the
$r$-bridge--between each pair of $P_{r-1},P_r$. Thus, we obtain
the sequence
\[ P_1, \beta_2, P_2, \beta_3,\dots, \beta_{2m-1}, P_{2m-1}.\]
Note that if any two permutations $\pi_1,\pi_2$ in the resulting
sequence satisfy $\pi_1 = (q,2m+2)\circ\pi_2$ for some $q\in[2m+1]$,
then w.l.o.g $\pi_2$ is odd and hence an $r$-bridge for some $r$,
and $\pi_1$ is even and thus not a bridge. Since in every bridge 
the last element is 
\[\beta_r(2m+2) = \phat_r(1) \in\mathset{1,2m+1},\]
and in every non-bridge it is not, it must follow, then, that 
$q=\beta_r(2m+2)$, and in particular
\begin{align*}
  \pi_1(2m+2) &= \parenv{\beta_r(2m+2),2m+2}\circ \beta_r(2m+2) \\
  &= 2m+2,
\end{align*}
thus $\pi_1\in P_0$.

We witness, therefore, that no such pair of permutations exist, since
we have not yet incorporated $P_0$ into our code. It also becomes
apparent that $P_0$ must necessarily be amended prior to its
inclusion, so it does not include any permutations of the form
\[\parenv{\beta_r(2m+2),2m+2}\circ \beta_r, \qquad 0<r\leq 2m.\]

In order to do so, we note that for all $r\geq 0$ 
\[\beta_r(2m) = \phat_r(2m+1) \in \mathset{1,2m+1},\]
and in particular $\beta_r(2m)\neq 2m+2$, hence 
\[\parenv{\beta_r(2m+2),2m+2}\circ \beta_r(2m) = \beta_r(2m) 
\in \mathset{1,2m+1}.\]

It follows that if we let $P_0^\prime$ be generated by the transition 
sequence $t_{\uparrow 2m-1}{}^{2m-1}$ applied to $\phat_0$, then it is 
parity-preserving, its last permutation is $t_{\uparrow 2m-1}{}^{-1} 
\phat_0$, and for all $\pi\in P_0^\prime$ we have $\pi(2m) = 2m 
\not\in\mathset{1,2m+1}$, thus
\[P_0^\prime \cap \mathset{\parenv{\beta_r(2m+2),2m+2}\circ 
\beta_r}_{r=1}^{2m} = \emptyset.\]

\begin{lemma}
  \label{aux:almost}
  The following sequence $P$,
  \[ P = P_0^\prime, \beta_1, P_1, \beta_2, P_2, \beta_3, \dots, 
  \beta_{2m-1},P_{2m-1},\beta_{2m},\]
  is a cyclic $\Gaux(2m+2,M)$.
\end{lemma}
\begin{IEEEproof}
By \autoref{aux:stitch}, and since when considered as sets,
\[ P_r \cap P_{r'} = \emptyset\]
for all $0 < r < r' < 2m$, and similarly $P_0^\prime$ is disjoint 
from $P_1, P_2,\ldots, P_{2m-1}$, we know that $P$ is a 
$G_\uparrow(2m+2, M)$.

As seen above, if for any two permutations $\pi_1,\pi_2\in P$ and 
$q\in[2m+1]$ we have $\pi_1 = (q,2m+2)\circ\pi_2$, then w.l.o.g. 
$\pi_2 = \beta_r$ for some $0<r\leq 2m$ and $q = \beta_r(2m+2)$. In 
particular, $\pi_1(2m+2) = 2m+2$, thus
\[\pi_1\not\in \mathset{\beta_r}_{r=1}^{2m}\cup
\bigcup_{r=1}^{2m-1} P_r,\]
thus $\pi_1\in P_0^\prime$. But 
\[P_0^\prime \cap \mathset{\parenv{\beta_r(2m+2),2m+2}\circ 
\beta_r}_{r=1}^{2m} = \emptyset,\]
in contradiction.
\end{IEEEproof}

The auxiliary code from \autoref{aux:almost} is almost what we
need. The only property lacking is the fact that $\id$ is not followed
in $P$ by the transition $t_{\uparrow 2m+2}$. We fix this in the
following theorem.

\begin{theorem}
  \label{aux:final}
  Let $k\geq 6$ be even. Then there exists a $\Gaux(k,M)$ starting at 
  $\id$ and a $t_{\uparrow k}$ transition, with
  \[M = \begin{cases}
  178 & k=6,\\
  (k-3)\parenv{\frac{(k-1)!}{2}+2}+1 & k>6.
  \end{cases}\]
  In particular, for all $k>6$, 
  \[M > \frac{k-3}{k}\cdot\frac{k!}{2}.\]
\end{theorem}
\begin{IEEEproof}
Denote $k=2m+2$ for $m\geq 2$, and let $P=(c_j)_{j=1}^M$ be the code 
from \autoref{aux:almost}. Since $\id\in S_k$ is not followed with 
a $t_{\uparrow k}$ transition in $P$, we denote the last permutation 
of $P_0^\prime$ by $\tilde{\pi}$, and replace $P$ with
\[\tilde{P} = \tilde{\pi}^{-1}P = \parenv{\tilde{\pi}^{-1}\circ c_j}_{j=1}^M.\]

We observe that $\tilde{P}$ is still a \quo{push-to-the-top} code
since \quo{push-to-the-top} transitions are group actions by
right-multiplications. Moreover, since $\tilde{\pi}(k)=k$, if for
some $\pi_1,\pi_2\in P$ we have $\tilde{\pi}^{-1}\circ\pi_1 =
(q,k)\circ\parenv{\tilde{\pi}^{-1}\circ\pi_2}$, where $q\in [k-1]$, then
\begin{align*}
\pi_1 &= \tilde{\pi}\circ\sparenv{(q,k)\circ\parenv{\tilde{\pi}^{-1}
\circ\pi_2}} \\
&= \sparenv{\tilde{\pi}\circ(q,k)\circ\tilde{\pi}^{-1}}\circ\pi_2 = 
\parenv{\tilde{\pi}(q), k}\circ\pi_2,
\end{align*}
and $\tilde{\pi}(q)\in [k-1]$, in contradiction.

As for the size of the code, note that $\abs{P_0^\prime} = 2m-1 = k-3$ and
\[\abs{P_1} = \abs{P_2} = \ldots = \abs{P_{2m-1}} = \begin{cases}
57 & k=6,\\
\frac{(k-1)!}{2} & k>6.
\end{cases}\]
Counting $\beta_1,\ldots,\beta_{2m}$, the claim is thus substantiated.
\end{IEEEproof}

To conclude this section, we combine \autoref{holsnakes}, \autoref{flipconst} 
and \autoref{aux:final} into the following corollary.

\begin{corollary}\label{auxcodes}
For all $k\geq 3$ there exists a $\Gaux(k,\tilde{M}_k)$ starting with
$\id$ and a $t_{\uparrow k}$ transition, where
\[\tilde{M}_k = \begin{cases}
8 & k=4;\\
57 & k=5;\\
178 & k=6;\\
\frac{k!}{2} & 5\neq k\equiv 1\pmod{2};\\
\rho_k\frac{k!}{2} & 6 < k\equiv 0\pmod{2},
\end{cases}\]
where $\rho_k > \frac{k-3}{k}$.
\end{corollary}

\section{Code Construction}\label{sec:const}

In this section we present the main construction of our paper, and 
discuss the size and asymptotic rate of the resulting codes. We will 
show, surprisingly, that our method generates codes which are larger 
than formerly known families of codes, even though we require the 
additional structure of a Gray code.

\subsection{Main code construction}
We now present a construction of $G_\uparrow(n,M,d)$ codes, for 
$d\leq n$, which we base on \autoref{auxcodes} and \autoref{jsnakes}.

It will simplify the presentation to assume $n=kd$ for some positive
$k\geq 2$, since in that case every congruence class modulo $d$ of
$[n]$ has size $k$. Nonetheless, the construction is applicable to any
$n>d$ with natural amendments. We discuss these changes, focusing on
special cases, after presenting the simple construction first.

\begin{construction}
  \label{con:main}
  Let $n,k,d\in\N$, with $n=kd$ and $k\geq 2$. We recursively
  construct a sequence of codes, $C_{d},C_{d-1},\dots,C_1$. An
  explicit construction is given for $C_{d}$ and a recursion step
  constructs $C_{m}$ from $C_{m+1}$.

  Recursion base: We construct the code $C_{d}$ by starting at the
  permutation $\sigma_0\in S_n$ defined by
  \[\sigma_0(j) = d\parenv{j \bmod k} + \ceilenv{\frac{j}{k}}.\]
  We obtain a transition sequence $t_{\uparrow r_1}, t_{\uparrow r_2}, 
  \ldots,t_{\uparrow r_{k!}}$ which generates the $G_\uparrow(k,k!)$ 
  provided by \autoref{jsnakes}. The code $C_{d}$ starts with 
  $\sigma_0$, and uses the transition sequence 
  \begin{multline*}
    t_{k(d-1)+1\uparrow k(d-1)+r_1},t_{k(d-1)+1\uparrow k(d-1)+r_2},\ldots\\
    \ldots,t_{k(d-1)+1\uparrow k(d-1)+r_{k!}}.
  \end{multline*}

  Recursion step: Assume $C_{m+1}$ has already been constructed, starting
  with permutation $\sigma_0$. Additionally, let
  \begin{equation}
    \label{eq:auxseq}
    t_{\uparrow k+1},t_{\uparrow i_2},\ldots,t_{\uparrow i_{\tilde{M}_{k+1}}}
  \end{equation}
  be a transition sequence generating a
  $\Gaux(k+1,\tilde{M}_{k+1})$ code provided by
  \autoref{auxcodes}.

  We construct the code $C_{m}$ as follows: replace each
  $t_{km+1\uparrow j}$ transition of $C_{m+1}$ with $t_{k(m-1)+1\uparrow
    j}$, followed by $t_{k(m-1)+1\uparrow k(m-1)+i_2}$,
  $t_{k(m-1)+1\uparrow k(m-1)+i_3}$, and so on until
  $t_{k(m-1)+1\uparrow k(m-1)+i_{\tilde{M}_{k+1}}}$.
\end{construction}

\begin{lemma}
  \label{lem:recurbase}
  For all $n=kd$, $k\geq 2$, the code $C_{d}$ from \autoref{con:main} is
  a $G_{k(d-1)+1\uparrow}(n,k!,d)$.
\end{lemma}
\begin{IEEEproof}
The parameters of the code are obvious, except perhaps the minimal
distance $d$. The fact that the codewords of $C_{d}$ are distinct
follows from \autoref{jsnakes}.
  
To prove the minimal distance $d$, note that for all $0\leq u<d$ and
$ku+1\leq i<j\leq k(u+1)$ it holds that
$\sigma_0(i)\equiv\sigma_0(j)\pmod d$. Thus, for every distinct
$\sigma,\tau\in C_{d}$, there exists $j$, $m(d-1)<j\leq kd=n$, such
that $\sigma(j)\neq \tau(j)$. Since by construction
$\sigma(j)\equiv\tau(j)\equiv 0\pmod d$, we observe
\[d_\infty(\sigma,\tau) \geq \abs{\sigma(j)-\tau(j)} \geq d,\]
implying that $C_{d}$ is a $G_{k(d-1)+1\uparrow}(n,k!,d)$.
\end{IEEEproof}

\begin{example}\label{exm:con:base}
We let $d=3$, $k=2$, and $n=kd=6$. We construct the code $C_3$
starting at
\[\sigma_0 = \sparenv{4, 1, 5, 2, 6, 3}\in S_6.\]
We use the complete $G_\uparrow(2,2)$ shown in \autoref{exm:aux},
which is generated by the sequence $t_{\uparrow 2}, t_{\uparrow
  2}$. We arrive at a generating sequence $t_{5\uparrow 6},
t_{5\uparrow 6}$ for $C_3$. Hence, in our example
\[C_3 = \parenv{\sparenv{4, 1, 5, 2, 6, 3}, \sparenv{4, 1, 5, 2, 3, 6}},\]
which is readily seen to be a $G_{5\uparrow}(6,2,3)$ code.
\end{example}

\begin{theorem}\label{iter_const}
  For all $n=kd$, $k\geq 2$, the code $C_1$ from \autoref{con:main} is
  a $G_{\uparrow}(n,\tilde{M}_{k+1}{}^{d-1}\cdot k!,d)$.
\end{theorem}
\begin{IEEEproof}
To prove the claim we will prove by induction that $C_{m}$ from
\autoref{con:main}, for all $m\in [d]$, is a
${G_{k(m-1)+1\uparrow}(n,\tilde{M}_{k+1}{}^{d-m}\cdot k!,d)}$. 
The base case of $C_d$ was proved in \autoref{lem:recurbase}. 
Assume the claim holds for $C_{m+1}$ and we now prove it for $C_{m}$.

Recall \eqref{eq:auxseq} gives the sequence of transitions for
a $\Gaux({k+1},\tilde{M}_{k+1})$. Then
\[
t_{\uparrow i_{\tilde{M}_{k+1}}}t_{\uparrow i_{\tilde{M}_{k+1}-1}}
\cdots t_{\uparrow i_3}t_{\uparrow i_2} = t_{\uparrow k+1}{}^{-1}.
\]
Thus,
\[t_{km+1\uparrow j} = \parenv{\prod_{r=2}^{\tilde{M}_{k+1}} 
t_{k(m-1)+1\uparrow k(m-1)+i_r}}t_{k(m-1)+1\uparrow j}\]
(where the product is expanded right-to-left). Therefore, $C_{m}$ 
expands each \quo{push-to-the-$km+1$st-index} transition of $C_{m+1}$ 
into $\tilde{M}_{k+1}$ \quo{push-to-the-$k(m-1)+1$st-index} transitions.

It follows that $C_{m}$ contains the codewords of $C_{m+1}$ in the same
order, with $\tilde{M}_{k+1}-1$ new words inserted between any two
words originally from $C_{m+1}$.  We say that each codeword of $C_{m+1}$
(now appearing in $C_{m}$) is the \emph{$C_{m+1}$-parent} of each of the
$\tilde{M}_{k+1}$ preceding codewords in $C_{m}$ (including itself),
since their vector notations agree on the order of the elements
\[\sigma_0(km+1), \sigma_0(km+2), \ldots, \sigma_0(n).\]

Now, suppose that $\sigma,\tau\in C_{m}$ satisfy
$d_\infty(\sigma,\tau)<d$. Let $\sigma^\prime$, $\tau^\prime$ be their
$C_{m+1}$-parents, respectively. To complete the proof we will show that
$\sigma=\tau$.

\textbf{Case 1:} $\sigma^\prime=\tau^\prime$. Denote 
\[x=\sigma^\prime(km+1) = \tau^\prime(km+1)\]
and $s = \sigma^{-1}(x)$, $a = \tau(s)$.

If $a = x$ then for all $j\neq s$, $k(m-1)<j\leq km+1$, we have 
$\sigma(j)\equiv\tau(j)\pmod d$ and
\[\abs{\sigma(j)-\tau(j)}\leq d_\infty(\sigma,\tau)<d,\]
hence $\sigma(j)=\tau(j)$, and
$\sigma=\tau$.

Otherwise, $a\neq x$, and denote $t = \tau^{-1}(x)\neq s$. It
similarly holds for all $j\not\in\mathset{s,t}$, $k(m-1)<j\leq
km+1$, that $\sigma(j)=\tau(j)$. We therefore observe $\tau =
\sigma\circ(s,t)$.  This implies that, if we let $\hat{\sigma},
\hat{\tau}\in S_{k+1}$ be the permutations in the $\Gaux$ we
obtained, generated similarly to $\sigma, \tau$, respectively (i.e., by
their corresponding transition sequences), then
\[\hat{\tau} = \hat{\sigma}\circ\parenv{s-k(m-1),t-k(m-1)} = 
(q,k+1)\hat{\sigma}\]
for some $q\in [k]$, in contradiction to the fact it was a 
${\Gaux(k+1,\tilde{M}_{k+1})}$.

\textbf{Case 2:} $\sigma^\prime\neq\tau^\prime$. Since $\sigma^\prime,
\tau^\prime\in C_{m+1}$ we have by assumption 
$d_\infty(\sigma^\prime,\tau^\prime)\geq d$, and note that for all $j$ 
satisfying ${j\leq k(m-1)}$ or $j>km+1$, it holds that
$\sigma(j)=\sigma^\prime(j)$ and $\tau(j) = \tau^\prime(j)$. Hence 
there exists $j$, $k(m-1)<j\leq km+1$, such that
\[\abs{\sigma(j)-\tau(j)}<d \quad \text{but}\quad \abs{\sigma^\prime(j)-\tau^\prime(j)}\geq d.\]
Note particularly, since for all $k(m-1)<j\leq km$ it holds that 
$\sigma^\prime(j) = \sigma_0(j) = \tau^\prime(j)$, that we have 
\[\abs{\sigma^\prime(km+1)-\tau^\prime(km+1)}\geq d.\]

Denote $x = \sigma^\prime(km+1)$, $y = \tau^\prime(km+1)$, and 
note that
\begin{align*}
\mathset{\sigma(j)}_{j=k(m-1)+1}^{km+1} &= \mathset{a_i}_{i=1}^{k}\cup\mathset{x};\\
\mathset{\tau(j)}_{j=k(m-1)+1}^{km+1} &= \mathset{a_i}_{i=1}^{k}\cup\mathset{y},
\end{align*}
where $\mathset{a_i}_{i=1}^{k}$ is a congruence class modulo $d$ of
$[n]$, of which $x,y$ are not members.

Let $s = \sigma^{-1}(x)$ and denote $a = \tau(s)$. Since 
\[\abs{x-a} = \abs{\sigma(s)-\tau(s)}\leq d_\infty(\sigma,\tau)<d\]
we have $a\neq y$. Let $t = \sigma^{-1}(a)$. Since 
$a\in\mathset{a_i}_{i=1}^{k}$ is a congruence class modulo $d$, 
for all $b\in\mathset{a_i}_{i=1}^{k}\setminus\mathset{a}$ we 
observe $\abs{a-b}\geq d$, but 
\[\abs{a-\tau(t)} = \abs{\sigma(t)-\tau(t)}\leq d_\infty(\sigma,\tau)<d\]
and therefore $\tau(t) = y$. For all $j\not\in\mathset{s,t}$ satisfying 
${k(m-1)}<j\leq km+1$ we then have ${\sigma(j)\equiv\tau(j)\pmod d}$ 
and $\abs{\sigma(j)-\tau(j)}\leq d_\infty(\sigma,\tau)<d$, hence 
$\sigma(j)=\tau(j)$.

This implies that, if we again let $\hat{\sigma}, \hat{\tau}\in S_{k+1}$ 
be the permutations in the $\Gaux$ generated similarly 
to $\sigma, \tau$ respectively, then 
\[\hat{\tau} = \hat{\sigma}\circ\parenv{s-k(m-1),t-k(m-1)} = 
(q,k+1)\hat{\sigma}\]
where $q$ is given by $a = a_q \in \mathset{a_i}_{i=1}^{k}$, again 
contradicting the properties of a ${\Gaux(k+1,\tilde{M}_{k+1})}$. Hence
$C_{m}$ has minimal $\ell_\infty$-distance of at least $d$, as required.
\end{IEEEproof}

\begin{example}\label{exm:con:rec}
We complete \autoref{exm:con:base} into a $G_\uparrow(6, 3^2\cdot 2, 3)$ 
code by applying the recursion step twice. In each step, since $k=2$, 
we utilize the trivial parity-preserving $\Gaux(3,3)$ code generated 
by the sequence $t_{\uparrow 3}, t_{\uparrow 3}, t_{\uparrow 3}$.

Firstly, recall that we used
\[\sigma_0 = \sparenv{4, 1, 5, 2, 6, 3}\in S_6,\]
and the sequence $t_{5\uparrow 6}, t_{5\uparrow 6}$ generates
\[C_3 = \parenv{\sparenv{4, 1, 5, 2, 6, 3}, \sparenv{4, 1, 5, 2, 3, 6}}.\]

We build $C_2$ by exchanging each $t_{5\uparrow 6}$ transition by 
$t_{3\uparrow 6}$ followed by 2 instances of $t_{2+1\uparrow 2+3} = 
t_{3\uparrow 5}$; the middle level of \autoref{fig:exm:con:rec} shows 
the resulting code.

Secondly, as seen in the same figure, each $t_{3\uparrow j}$ transition 
of $C_2$, $j\in\mathset{5,6}$, can be replaced by $t_{1\uparrow j} = 
t_{\uparrow j}$, followed by 2 instances of $t_{0+1\uparrow 0+3} = 
t_{\uparrow 3}$, to generate $C_1$.

Note that $C_3\subseteq C_2\subseteq C_1$, and that they are 
$G_{5\uparrow}(6,2,3)$, $G_{3\uparrow}(6,6,3)$ and 
$G_{\uparrow}(6,18,3)$ codes, respectively.
\end{example}
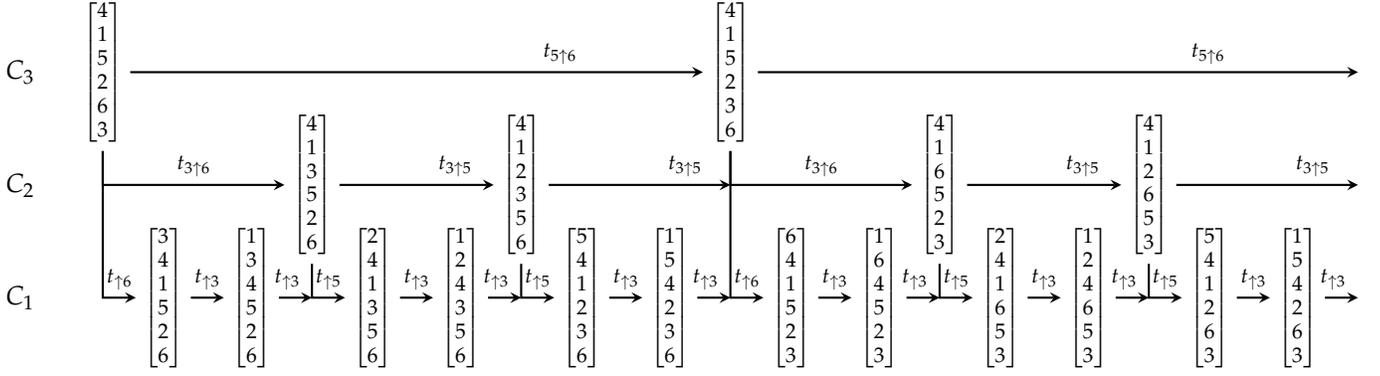
\begin{figure*}[ht]
\begin{center}
\begin{footnotesize}
\begin{tikzpicture}[node distance=15 mm and 4.4mm]
\tikzstyle{perm} = [draw=none,fill=none]
\tikzstyle{arrow} = [thick,->,>=stealth]

\node (c3) [perm] {\begin{normalsize}$C_3$\end{normalsize}};
\node (c3-0) [perm, right=of c3] {$\begin{bmatrix}4\\ 1\\ 5\\ 2\\ 6\\ 3\end{bmatrix}$};

\node (c2) [perm, below of=c3] {\begin{normalsize}$C_2$\end{normalsize}};
\coordinate[below of=c3-0] (c2-0_0);

\node (c1) [perm, below of=c2] {\begin{normalsize}$C_1$\end{normalsize}};
\coordinate[below of=c2-0_0] (c1-0_0_0);
\node (c1-0_0_1) [perm, right=of c1-0_0_0] {$\begin{bmatrix}3\\ 4\\ 1\\ 5\\ 2\\ 6\end{bmatrix}$};
\node (c1-0_0_2) [perm, right=of c1-0_0_1] {$\begin{bmatrix}1\\ 3\\ 4\\ 5\\ 2\\ 6\end{bmatrix}$};
\coordinate[right=of c1-0_0_2] (c1-0_1_0);
\node (c1-0_1_1) [perm, right=of c1-0_1_0] {$\begin{bmatrix}2\\ 4\\ 1\\ 3\\ 5\\ 6\end{bmatrix}$};
\node (c1-0_1_2) [perm, right=of c1-0_1_1] {$\begin{bmatrix}1\\ 2\\ 4\\ 3\\ 5\\ 6\end{bmatrix}$};
\coordinate[right=of c1-0_1_2] (c1-0_2_0);
\node (c1-0_2_1) [perm, right=of c1-0_2_0] {$\begin{bmatrix}5\\ 4\\ 1\\ 2\\ 3\\ 6\end{bmatrix}$};
\node (c1-0_2_2) [perm, right=of c1-0_2_1] {$\begin{bmatrix}1\\ 5\\ 4\\ 2\\ 3\\ 6\end{bmatrix}$};
\coordinate[right=of c1-0_2_2] (c1-1_0_0);
\node (c1-1_0_1) [perm, right=of c1-1_0_0] {$\begin{bmatrix}6\\ 4\\ 1\\ 5\\ 2\\ 3\end{bmatrix}$};
\node (c1-1_0_2) [perm, right=of c1-1_0_1] {$\begin{bmatrix}1\\ 6\\ 4\\ 5\\ 2\\ 3\end{bmatrix}$};
\coordinate[right=of c1-1_0_2] (c1-1_1_0);
\node (c1-1_1_1) [perm, right=of c1-1_1_0] {$\begin{bmatrix}2\\ 4\\ 1\\ 6\\ 5\\ 3\end{bmatrix}$};
\node (c1-1_1_2) [perm, right=of c1-1_1_1] {$\begin{bmatrix}1\\ 2\\ 4\\ 6\\ 5\\ 3\end{bmatrix}$};
\coordinate[right=of c1-1_1_2] (c1-1_2_0);
\node (c1-1_2_1) [perm, right=of c1-1_2_0] {$\begin{bmatrix}5\\ 4\\ 1\\ 2\\ 6\\ 3\end{bmatrix}$};
\node (c1-1_2_2) [perm, right=of c1-1_2_1] {$\begin{bmatrix}1\\ 5\\ 4\\ 2\\ 6\\ 3\end{bmatrix}$};
\coordinate[right=of c1-1_2_2] (c1-2_0_0);

\node (c2-0_1) [perm, above of=c1-0_1_0] {$\begin{bmatrix}4\\ 1\\ 3\\ 5\\ 2\\ 6\end{bmatrix}$};
\node (c2-0_2) [perm, above of=c1-0_2_0] {$\begin{bmatrix}4\\ 1\\ 2\\ 3\\ 5\\ 6\end{bmatrix}$};
\coordinate[above of=c1-1_0_0] (c2-1_0);
\node (c2-1_1) [perm, above of=c1-1_1_0] {$\begin{bmatrix}4\\ 1\\ 6\\ 5\\ 2\\ 3\end{bmatrix}$};
\node (c2-1_2) [perm, above of=c1-1_2_0] {$\begin{bmatrix}4\\ 1\\ 2\\ 6\\ 5\\ 3\end{bmatrix}$};
\coordinate[above of=c1-2_0_0] (c2-2_0);

\node (c3-1) [perm, above of=c2-1_0] {$\begin{bmatrix}4\\ 1\\ 5\\ 2\\ 3\\ 6\end{bmatrix}$};
\coordinate[above of=c2-2_0] (c3-2);

\draw [arrow] (c3-0) -- (c3-1) node[near end,above] {$t_{5 \uparrow 6}$};
\draw [arrow] (c3-0) |- (c2-0_1) node[near end,above] {$t_{3 \uparrow 6}$};
\draw [arrow] (c3-0) |- (c1-0_0_1) node[near end,above] {$t_{\uparrow 6}$};
\draw [arrow] (c3-1) -- (c3-2) node[near end,above] {$t_{5 \uparrow 6}$};
\draw [arrow] (c3-1) |- (c2-1_1) node[near end,above] {$t_{3 \uparrow 6}$};
\draw [arrow] (c3-1) |- (c1-1_0_1) node[near end,above] {$t_{\uparrow 6}$};

\draw [arrow] (c2-0_1) -- (c2-0_2) node[near end,above] {$t_{3 \uparrow 5}$};
\draw [arrow] (c2-0_1) |- (c1-0_1_1) node[near end,above] {$t_{\uparrow 5}$};
\draw [arrow] (c2-0_2) -- (c2-1_0) node[near end,above] {$t_{3 \uparrow 5}$};
\draw [arrow] (c2-0_2) |- (c1-0_2_1) node[near end,above] {$t_{\uparrow 5}$};

\draw [arrow] (c2-1_1) -- (c2-1_2) node[near end,above] {$t_{3 \uparrow 5}$};
\draw [arrow] (c2-1_1) |- (c1-1_1_1) node[near end,above] {$t_{\uparrow 5}$};
\draw [arrow] (c2-1_2) -- (c2-2_0) node[near end,above] {$t_{3 \uparrow 5}$};
\draw [arrow] (c2-1_2) |- (c1-1_2_1) node[near end,above] {$t_{\uparrow 5}$};

\draw [arrow] (c1-0_0_1) -- (c1-0_0_2) node[midway,above] {$t_{\uparrow 3}$};
\draw [arrow] (c1-0_0_2) -- (c1-0_1_0) node[near start,above] {$t_{\uparrow 3}$};

\draw [arrow] (c1-0_1_1) -- (c1-0_1_2) node[midway,above] {$t_{\uparrow 3}$};
\draw [arrow] (c1-0_1_2) -- (c1-0_2_0) node[near start,above] {$t_{\uparrow 3}$};

\draw [arrow] (c1-0_2_1) -- (c1-0_2_2) node[midway,above] {$t_{\uparrow 3}$};
\draw [arrow] (c1-0_2_2) -- (c1-1_0_0) node[near start,above] {$t_{\uparrow 3}$};

\draw [arrow] (c1-1_0_1) -- (c1-1_0_2) node[midway,above] {$t_{\uparrow 3}$};
\draw [arrow] (c1-1_0_2) -- (c1-1_1_0) node[near start,above] {$t_{\uparrow 3}$};

\draw [arrow] (c1-1_1_1) -- (c1-1_1_2) node[midway,above] {$t_{\uparrow 3}$};
\draw [arrow] (c1-1_1_2) -- (c1-1_2_0) node[near start,above] {$t_{\uparrow 3}$};

\draw [arrow] (c1-1_2_1) -- (c1-1_2_2) node[midway,above] {$t_{\uparrow 3}$};
\draw [arrow] (c1-1_2_2) -- (c1-2_0_0) node[near start,above] {$t_{\uparrow 3}$};
\end{tikzpicture}
\end{footnotesize}
\end{center}\vspace*{-2ex}
\caption{\autoref{con:main} as demonstrated in the case $d=3$, $k=2$.
\label{fig:exm:con:rec}}
\end{figure*}

We now describe the changes needed in \autoref{con:main} to allow
general $n$ and $d$ parameters. We first consider $n$ not necessarily
being a multiple of $d$, but still $n\geq 2d$. For all $i\in[d]$, let
\[ \cR_i = \mathset{ i, i+d, i+2d, \dots, n-\parenv{(n-i)\bmod d }},\]
be the $i$th congruence class modulo $d$ of $[n]$. Then
\[\abs{\cR_i} = \begin{cases}
  \ceilenv{\frac{n}{d}} & 1\leq i\leq (n\bmod d),\\
  \floorenv{\frac{n}{d}} & (n\bmod d) < i\leq d.
\end{cases}
\]
We define the starting permutation
\[ \sigma_0 = [ \cR_1 | \cR_2 | \dots | \cR_d ]\in S_n,\]
to be comprised of a concatenation of the congruence classes, where
the order of elements within the congruence class is
arbitrary. Additionally, the recursion base uses a
$G_\uparrow(\abs{\cR_d},\abs{\cR_d}!)$. As for the recursion step of
constructing $C_{m}$ from $C_{m+1}$, we can still apply it with the
following changes:
\begin{itemize}
\item
  We choose $\Gaux(\abs{\cR_{m}}+1,\tilde{M}_{\abs{\cR_{m}}+1})$.
\item
  We use push operations to position $1+\sum_{i=1}^{m-1} \abs{\cR_i}$.
\end{itemize}
We obtain $C_1$ which is a $G_{\uparrow}\parenv{n,M,d}$, where
\[M = \tilde{M}_{\ceilenv{n/d}+1}{}^{n\bmod d}\cdot 
\tilde{M}_{\floorenv{n/d}+1}{}^{d-(n\bmod d)-1}\cdot
\floorenv{\frac{n}{d}}!.\]

Finally, we discuss the special case of $n<2d$, in which all but
$\parenv{n\bmod d}$ congruence classes are singletons. We will amend
our construction by replacing the recursion base with
\[C_{m} = \mathset{\sigma_0,t_{2m-1\uparrow 2m+1}\sigma_0,
  t_{2m-1\uparrow 2m+1}{}^2\sigma_0},\]
where $m = n\bmod d$, and
continuing the recursion step as discussed above.  Thus, we are
effectively only using the first member of $\cR_{m+1}$ together with the
previous congruence classes, fixing $\sigma_0(j)$ for $j>2m+1$. In this
case, we obtain $C_1$ which is a $G_{\uparrow}(n,3^{n\bmod d},d)$.

Thus, in what follows, whenever we mention \autoref{con:main}, we
refer to its most general version applying to all $n$ and $d$.

\subsection{Code-size analysis and comparison}

We would like to give an explicit expression for the size of the codes
constructed by \autoref{con:main}. This would enable a comparison with
previously known results.

\begin{lemma}
  Let $C_1$ be the code from (the general version of)
  \autoref{con:main}.  Then its size, $\abs{C_1}$, is given by
  \eqref{eq:codesize}.
\end{lemma}
\begin{IEEEproof}
Let us first assume $n\geq 2d$. We note the asymmetry in
\autoref{con:main} between congruence classes $\cR_i$ of odd and even
sizes. Indeed, a class of size $\abs{\cR_i}=k\geq 2$ (for all classes
other than $\cR_d$, which is used in the recursion base and whose
contribution is based on the $G_\uparrow(k,k!)$ code) contributes to
the code size, according to \autoref{auxcodes}, a multiplicative factor of
\[\tilde{M}_{k+1} = \begin{cases}
8 & k=3;\\
57 & k=4;\\
178 & k=5;\\
\frac{(k+1)!}{2} & 4 \neq k\equiv 0\pmod{2};\\
\rho_{k+1}\frac{(k+1)!}{2} & 5 < k\equiv 1\pmod{2},
\end{cases}\]
where, again, $\rho_{k+1} > \frac{k-2}{k+1}$.

\begin{figure*}
  \hrulefill
  \begin{equation}
    \label{eq:codesize}
    \abs{C_1}= \begin{cases}
      \parenv{\ceilenv{\frac{n}{d}}+1}^{n\bmod d}
      \left.\parenv{\floorenv{\frac{n}{d}}+1}!\right.^d
      \cdot\frac{\rho_{\ceilenv{n/d}+1}{}^{n\bmod d}}{2^{d-1}\parenv{\floorenv{n/d}+1}}
        & 4 < \floorenv{\frac{n}{d}}\equiv 0 \pmod{2},\\
      \parenv{\frac{178}{57}}^{n\bmod d}
      \cdot 57^{d-1}
      \cdot 24 & \floorenv{\frac{n}{d}}=4,\\
      \parenv{\frac{8}{3}}^{n\bmod d}\cdot 3^{d-1}\cdot 2
         & \floorenv{\frac{n}{d}}=2,\\
      \parenv{\ceilenv{\frac{n}{d}}+1}^{n\bmod d}
      \left.\parenv{\floorenv{\frac{n}{d}}+1}!\right.^d
      \cdot\frac{\rho_{\floorenv{n/d}+1}{}^{(d-1)-(n\bmod d)}}{2^{d-1}\parenv{\floorenv{n/d}+1}}
        & 5<\floorenv{\frac{n}{d}}\equiv 1 \pmod{2},\\
      \parenv{\frac{1260}{89}}^{n\bmod d}\cdot 
      \parenv{178}^d \cdot 
      \frac{120}{178} & \floorenv{\frac{n}{d}} = 5,\\
      \parenv{\frac{57}{8}}^{n\bmod d}\cdot 8^d
      \cdot\frac{3}{4} & \floorenv{\frac{n}{d}}=3,\\
      3^{n\bmod d} & \floorenv{\frac{n}{d}}=1.
    \end{cases}
  \end{equation}
  \hrulefill
\end{figure*}

It is therefore important to note that when
$\floorenv{\frac{n}{d}}\equiv 0\pmod 2$, $[n]$ has $\parenv{n\bmod d}$
congruence classes modulo $d$ of odd size $\ceilenv{\frac{n}{d}}$, and
$d-\parenv{n\bmod d}$ classes of even size $\floorenv{\frac{n}{d}}$.
Thus, if additionally $\floorenv{\frac{n}{d}} > 4$, the constructed code
$C_1$ is of size
\begin{align*}
\abs{C_1} &= \parenv{\rho_{\ceilenv{n/d}+1}\frac{(\ceilenv{n/d}+1)!}{2}}^{n\bmod d}\cdot
\floorenv{\frac{n}{d}}! \nonumber \\
&\quad \cdot\parenv{\frac{(\floorenv{n/d}+1)!}{2}}^{d-(n\bmod d)-1},
\end{align*}
and simple rearranging gives us the first case of \eqref{eq:codesize}.
Similar considerations give us the next five cases of \eqref{eq:codesize}.

Finally, we consider the case of $n<2d$, which implies
$\floorenv{\frac{n}{d}}=1$. In this special case we only permute
$(n\bmod d)=(n-d)$ congruence classes of $[n]$, (and each such class
has $2=\floorenv{\frac{n}{d}}+1$ elements).  As mentioned, we
therefore construct a code of size $\abs{C_1} = 3^{n\bmod d}$.
\end{IEEEproof}

We comment that it is also possible to achieve a slight gain in code
size by reordering $\sigma_0$ so that the last block consists of a
congruence class of odd size, rather than even, where the added
complexity of index calculation is inconsequential. The asymptotic
gain in code rate vanishes.

We now turn to comparing the size of the resulting code with that of
previously constructed codes, as well as known bounds on the
cardinality of such codes.

The first comparison we make is with codes that have the Gray
property. Such codes were only studied for $d=2$, i.e.,
snake-in-the-box codes or $G_\uparrow(n,M,2)$ codes in our
notation. These codes were studied in \cite[Thm.~24]{YehSch12b},
where it was shown that such codes can be constructed with sizes
\[M = \ceilenv{\frac{n}{2}}!
\parenv{\floorenv{\frac{n}{2}} +
  \parenv{\floorenv{\frac{n}{2}}-1}!}.\]
\autoref{con:main} improves this size by a factor of
$\frac{1}{2}\parenv{\ceilenv{\frac{n}{2}}+1}\floorenv{\frac{n}{2}}$, 
times $\rho_{\floorenv{n/2}+1}$ when $n\equiv 2\pmod{4}$ (in the case of 
$n\equiv 1\pmod{4}$ $\rho_{\floorenv{n/2}+1}$ is eliminated by changing 
the order of congruence classes in $\sigma_0$). We note that a similar 
improvement was made concurrently by \cite{WanFu16} in a preprint devoted 
solely to the case of $d=2$, i.e., snake-in-the-box codes.

We now also compare our results to error-correcting codes with the
$\ell_\infty$-metric which are not necessarily Gray codes
(LMRM-codes). We observe that the best known general LMRM-code
construction to date, \cite[Cst.~1, Thm.~2]{TamSch10} and
\cite[Sec.~III-A]{KloLinTsaTze10}, presented $(n,M,d)$-LMRM codes with
sizes
\[M = \left.\ceilenv{\frac{n}{d}}!\right.^{n\bmod d}
\left.\floorenv{\frac{n}{d}}!\right.^{d-\parenv{n\bmod d}},\]
which our construction improves upon, more pronouncedly the more 
$[n]$ has even-sized congruence classes modulo $d$ 
(cf.~\eqref{eq:codesize}).

%% We also compare our code construction in the asymptotic regime. We
%% first briefly follow \cite{YehSch12b} in examining the case of fixed
%% $d$, as $n$ grows to infinity. For ease of calculation, we observe the
%% worst case $n=(2k+1)d$ for $k\to\infty$,
%% and we have 
%% \[\abs{C_1} = M = \parenv{\frac{n}{d}!}^d
%% \parenv{1+\frac{1}{n/d}}^{d-1} > \parenv{\frac{n}{d}!}^d\]
%% Note in particular, since $m\log\frac{m}{e}<\log(m!)<m\log m$, that
%% \begin{align*}
%% \frac{\log M}{\log n!} & > 
%% \frac{d\log((2k+1)!)}{\log(((2k+1)d)!)} \\
%% &> \frac{\log(2k+1) - \log(e)}{\log(2k+1)+\log(d)}
%% \tends{k\to\infty}{1}
%% \end{align*}

In the asymptotic regime, we go on to examine the case of
$d=\Theta(n)$. For an $(n,M,d)$-LMRM code (and in particular a
$G_\uparrow(n,M,d)$), we follow the convention (e.g., \cite{TamSch10})
of defining the \emph{rate} of the code
\[R=\frac{1}{n}\log_2 M,\]
and its \emph{normalized distance} 
\[\delta = \frac{d}{n}.\]
The following were proven in \cite{TamSch10}.

\begin{lemma}\label{bound:rate:up}\cite[Thm.~23]{TamSch10}
For any $(n,M,d)$-LMRM code it holds that 
\begin{align*}
R\leq &\ 2 - 2\delta\floorenv{\frac{1}{\delta}} - 
\parenv{\delta\floorenv{\frac{1}{\delta}} - \delta}\log_2(\delta)\\
&- \parenv{1+\delta-\delta\floorenv{\frac{1}{\delta}}}
\log_2\parenv{1+\delta-\delta\floorenv{\frac{1}{\delta}}} + o(1).
\end{align*}
\end{lemma}

\begin{lemma}\label{const:Tam}\cite[Thm.~27]{TamSch10}
For any $0<\delta\leq 1$ the construction of 
\cite[Cst.~1, Thm.~2]{TamSch10} and \cite[Sec.~III-A] {KloLinTsaTze10}
yields codes with
\begin{align*}
R =& \parenv{1 - \delta\floorenv{\frac{1}{\delta}}}
\log_2\parenv{\ceilenv{\frac{1}{\delta}}!}\\
&+ \parenv{\delta + \delta\floorenv{\frac{1}{\delta}} - 1}
\log_2\parenv{\floorenv{\frac{1}{\delta}}!}.
\end{align*}
\end{lemma}

It was also shown in \cite{TamSch10} that a Gilbert-Varshamov-like
non-constructive bound exists:
\begin{lemma}\label{bound:rate:low}\cite[Thm.~25]{TamSch10}
For any $0<\delta\leq 1$ there exist $(n,M,d)$-LMRM codes satisfying 
$\frac{d}{n}\geq\delta$ with rate 
$R\geq f_{\mathrm{GV}}(\delta) + o(1)$, where
\[
f_{\mathrm{GV}} = \begin{cases}
\log_2\frac{1}{\delta} + 2\delta\parenv{\log_2 e-1}-1 & 0<\delta\leq \frac{1}{2}\\
-2\delta\log_2\frac{1}{\delta} + 2(1-\delta)\log_2 e & \frac{1}{2}\leq \delta\leq 1
\end{cases}
\]
\end{lemma}

We therefore aim to show that our construction can bridge some of the 
gap between the given bounds and known constructions.

\begin{figure*}
  \hrulefill
  \begin{equation}
    \label{eq:coderate}
    R \geq \begin{cases}
      \parenv{1 - \delta\floorenv{\frac{1}{\delta}}}
      \log_2\parenv{\ceilenv{\frac{1}{\delta}}+1}
      + \delta\log_2\parenv{\parenv{\floorenv{\frac{1}{\delta}}+1}!}\\
      \qquad + \parenv{1-\delta\floorenv{\frac{1}{\delta}}}
      \log_2\parenv{\frac{\ceilenv{n/d}-2}{\ceilenv{n/d}+1}} - \delta
       & 4 < \floorenv{1/\delta}\equiv 0 \pmod{2},\\
      
      \parenv{1 - 4\delta}\parenv{\log_2(178)}
      + \parenv{5\delta-1}\log_2\parenv{57}
        & \floorenv{1/\delta}=4,\\
      
      \parenv{1 - 2\delta}\parenv{3-\log_2\parenv{3}}
      + \delta\log_2\parenv{3}
        & \floorenv{1/\delta}=2,\\
      
      \parenv{1 - \delta\floorenv{\frac{1}{\delta}}}
      \log_2\parenv{\ceilenv{\frac{1}{\delta}}+1}
      + \delta\log_2\parenv{\parenv{\floorenv{\frac{1}{\delta}}+1}!}\\
      \qquad + \parenv{\delta+\delta\floorenv{\frac{1}{\delta}}-1}
      \log_2\parenv{\frac{\floorenv{n/d}-2}{\floorenv{n/d}+1}} - \delta
        & 5 < \floorenv{1/\delta}\equiv 1 \pmod{2},\\
      
      \parenv{1-5\delta}\log_2\parenv{315}
      + \parenv{6\delta-1}\log_2(89) + 2 - 9\delta
        & \floorenv{1/\delta}=5,\\
      
      \parenv{1 - 3\delta}\parenv{\log_2\parenv{57}-4} + 1
        & \floorenv{1/\delta}=3,\\
      
      \parenv{1 - \delta}\log_2(3)
        & \floorenv{1/\delta}=1.
    \end{cases}
  \end{equation}
  \hrulefill
\end{figure*}

\begin{lemma}
  Let $C_1$ be the code from (the general version of) \autoref{con:main}. Then
  and estimate from below of its rate $R$ as a function of its normalized 
  distance $\delta$ is given by \eqref{eq:coderate}.
\end{lemma}
\begin{IEEEproof}
The proof follows by a simple substitution of 
$(n\bmod d) = n-d\floorenv{\frac{n}{d}}$ and $d=n\delta$ into 
\eqref{eq:codesize}. We also recall that $\rho_k>\frac{k-3}{k}$.
\end{IEEEproof}

\begin{figure*}[ht]
\psfrag{xax}{\small{(a)}}
\psfrag{xbx}{\small{(b)}}
\psfrag{xcx}{\small{(c)}}
\psfrag{xdx}{\small{(d)}}
\psfrag{rrr}{$R$}
\psfrag{delta}{$\delta$}
\vspace*{-20ex}
\includegraphics[width=\linewidth]{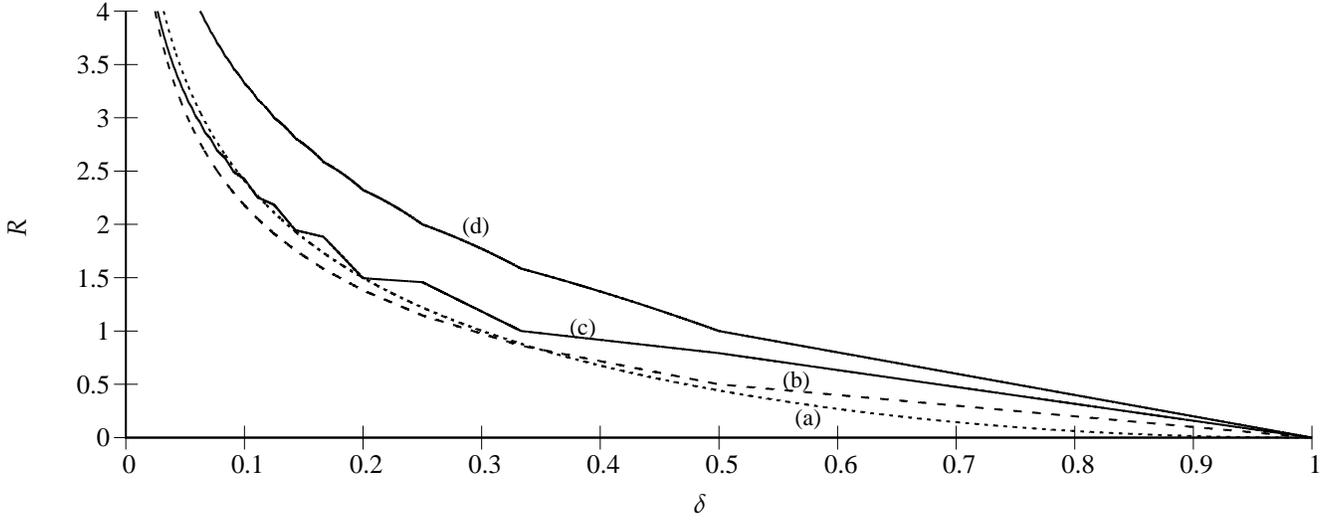}
\vspace*{-20ex}
\caption{(a) The Gilbert-Varshamov-like lower bound of
  \autoref{bound:rate:low}. (b) The rate of codes from
  \autoref{const:Tam} constructed in \cite{TamSch10}. (c) A lower bound 
  for the rate of codes $C_1$ from \autoref{con:main}. (d) The upper 
  bound of \autoref{bound:rate:up}.}
\label{figbounds}
\end{figure*}

In conclusion, these asymptotic rates and bounds are shown in
\autoref{figbounds}. We note in particular that the rate of codes
produced by \autoref{con:main} is strictly higher than that of
previously known constructions (as in \autoref{const:Tam}).
Furthermore, it produces codes with rates higher than those guaranteed
by the Gilbert-Varshamov-like \autoref{bound:rate:low} for all $\delta$ 
greater than $\approx 0.1$ except in a small neighborhood of 
$\frac{1}{5}$, whereas known constructions only bypassed these rates 
only for $\delta$ greater than $\approx 0.349$.

\section{Decoding Algorithm}
\label{sec:decodalg}

This section is devoted to devising a decoding algorithm capable of
correcting a noisy received version of a transmitted codeword.

Known constructions of $(n,M,d)$-LMRM codes, presented in
\cite[Cst.~1, Thm.~2]{TamSch10} and \cite[Sec.~III-A]{KloLinTsaTze10},
lend themselves to straightforward decoding algorithms, efficiently
done in $O(n)$ operations, since for any given codeword $\sigma$ and
index $i\in[n]$, $r=\sparenv{\sigma(i)\bmod d}$ is known. Hence, if a
retrieved permutation $\tau$ satisfies
$d_\infty(\sigma,\tau)\leq\floorenv{(d-1)/2}$, then $\sigma(i)$ is
known to be the unique element $k\in r+d\Z$ satisfying
$\abs{k-\tau(i)}\leq \floorenv{(d-1)/2}$.

Our proposed construction diverges from that rigid partition. However,
we can still efficiently decode noisy information, provided errors of
magnitude no more than $t$ have occurred, where $2t+1\leq d$. More
precisely, we assume that for every stored permutation $\sigma$ and
retrieved permutation $\tau$ it holds that $d_\infty(\sigma,\tau)\leq
t\leq \floorenv{(d-1)/2}$.

To simplify our presentation we assume $n=kd$, since then our
construction only makes (repeated) use of a single auxiliary
$\Gaux(k+1,\tilde{M}_{k+1})$ code. Extensions to the general
version of \autoref{con:main} are easily obtainable.

We first require a function $\mathtt{ValidAux}$ capable of detecting
whether a given permutation $\sigma\in S_{k+1}$ belongs to the
auxiliary $\Gaux(k+1,\tilde{M}_{k+1})$ code.

\begin{lemma}\label{validaux}
For an auxiliary $\Gaux(k+1,\tilde{M}_{k+1})$ code provided by 
\autoref{holsnakes}, \autoref{flipconst} or \autoref{aux:final}, 
a function $\mathtt{ValidAux}$ can be implemented to operate in 
$O(k)$ steps.
\end{lemma}
\begin{IEEEproof}
If we use \autoref{holsnakes}, then the auxiliary code consists of all
even permutations, and it is well known that we can determine the 
signature of a permutation $\sigma\in S_{k+1}$ in $O(k)$ operations, 
e.g., by finding a cycle decomposition of $\sigma$.

If we instead use \autoref{flipconst}, then the auxiliary code
consists of exactly those permutations in whose vector notation $1$
follows ${k+1}$ (cyclically), i.e.,
\[\sigma\parenv{(\sigma^{-1}(k+1)\bmod (k+1))+1}=1,\]
which again requires $O(k)$ steps to verify.

Finally, for \autoref{aux:final} we divide into cases according to 
$\sigma(k)$. For all elements other than $1,k-1,k$ the problem again 
reduces to determining $\operatorname*{sign}\sigma$. For those elements, 
only cyclic shift (on a subset of indices, by case) of a known 
permutation are valid, which we can easily verify in linear time.
\end{IEEEproof}

An important notion of a \emph{window} will be useful. Let $\sigma\in 
S_n$ be a permutation, $n=kd$. For all $j\in[d]$ we define the $j$th 
window as the set of indices
\[W_j = \mathset{ k(j-1)+2,k(j-1)+3,\dots,kj+1 }\cap[n].\]
The windows partition $[n]\setminus\mathset{1}$, and are all of size $k$
except $W_d$ which is of size $k-1$.

Given a set $I\subseteq[n]$, we conveniently denote
\[\sigma(I) = \mathset{\sigma(i) ~|~ i\in I }.\]
We prove a simple lemma concerning properties of windows of
codewords from \autoref{con:main}.

\begin{lemma}
  \label{lem:decodrecur}
  Let $\sigma$ be a codeword of $C_1$ from \autoref{con:main}, with
  $n=kd$. Then for all $j\in[d]$,
  \[ k-1 \leq \abs{\sigma(W_j)\cap \cR_j} \leq k,\]
  i.e., at most one element of $\sigma(W_j)$ does not leave a residue of 
  $j$ modulo $d$. In particular, $\sigma(W_d)\subseteq \cR_d$.

  Additionally, if $\abs{\sigma(W_j)\cap \cR_j}=k-1$, $j\in[d-1]$, then
  $x\in \sigma(W_j)\setminus\cR_j$ satisfies $x\in \cR_{j'}$ for
  some $j'> j$.
\end{lemma}
\begin{IEEEproof}
Take any $1<j\in[d]$, and let $\sigma_j$ be the $C_{j}$-parent of
$\sigma$. Then, in $C_1$, no transition between $\sigma$ and $\sigma_j$ 
is induced by $C_j$, and hence $\sigma_j$ is derived from $\sigma$ by a 
(perhaps empty) sequence of $t_{\uparrow i^\prime}$ transitions, for 
$i^\prime\in W_1\cup\dots \cup W_{j-1}$. Therefore, for all $i\in W_j 
\cup\dots \cup W_d$ we have $\sigma_j(i) = \sigma(i)$, and the same also 
holds for $j=1$ (since $\sigma_1=\sigma$). In particular, $\sigma(W_j) = 
\sigma_j(W_j)$.

Now, since $C_j$ only applies \quo{push-to-the-$k(j-1)+1$st-index} 
transitions, and
\[\sigma_0\parenv{\mathset{k(j-1)+1}\cup W_j\cup\dots \cup W_d} = 
\cR_j\cup\dots\cup\cR_d,\]
if for any $i\in W_j$ we have $\sigma(i) = \sigma_j(i)\not\in\cR_j$, 
then by necessity $\sigma(i)\in\cR_{j^\prime}$ for some $j^\prime>j$. 
In particular, $\sigma(W_d)\subseteq\cR_d$.

For all $j\in[d-1]$, we also consider $\sigma_{j+1}$, the 
$C_{j+1}$-parent of both $\sigma$ and $\sigma_j$. Since $C_{j+1}$ only 
applies \quo{push-to-the-$kj+1$st-index} transitions, 
\begin{align*}
\sigma_{j+1}&\parenv{(\mathset{k(j-1)+1}\cup W_j)
\setminus\mathset{kj+1}}\\
&= \sigma_0\parenv{(\mathset{k(j-1)+1}\cup W_j)
\setminus\mathset{kj+1}} = \cR_j.
\end{align*}

Finally, since $\sigma_{j+1}$ is derived from $\sigma_j$ 
by a sequence of $t_{k(j-1)+1\uparrow i^\prime}$ transitions for 
$i^\prime\in W_j$, it follows that 
\[\sigma_{j+1}\parenv{\mathset{k(j-1)+1}\cup W_j} = 
\sigma_j\parenv{\mathset{k(j-1)+1}\cup W_j}\]
thus
\begin{align*}
\sigma_j(W_j) &\subseteq 
\sigma_{j+1}\parenv{\mathset{k(j-1)+1}\cup W_j}\\
&= \cR_j\cup\mathset{\sigma_{j+1}(kj+1)}.
\end{align*}
Noting that $\abs{\sigma_j(W_j)}=\abs{\cR_j}=k$ and recalling 
that $\sigma(W_j) = \sigma_j(W_j)$, we are done.
\end{IEEEproof}

\begin{corollary}
  \label{cor:missing-element}
  Let $\sigma$ be a codeword of $C_1$ from \autoref{con:main}, with
  $n=kd$. Then for each $j\in[d]$, there is a unique element 
  $x^\sigma_j\in \cR_{j}\cup\dots\cup\cR_d$ satisfying 
  \[ \sigma(W_{j}\cup\dots\cup W_{d}) = 
  \cR_{j} \cup\dots\cup \cR_d \setminus\mathset{x^\sigma_j}.\]
\end{corollary}
\begin{IEEEproof}
The proposition follows from \autoref{lem:decodrecur} for $j=d$ since
$\abs{\sigma(W_d)} = \abs{W_d}=k-1 \leq\abs{\cR_d\cap\sigma(W_d)}$.
Now suppose the proposition holds for $j+1$, and we prove that
it holds for $j$.

We again observe by \autoref{lem:decodrecur} that $|\cR_{j}\cap
\sigma(W_{j})|\in\mathset{k-1,k}$.  If $|\cR_{j}\cap
\sigma(W_{j})|=k$, since $|\sigma(W_{j})| =
|W_{j}| = k$, then $\cR_{j} = \sigma(W_{j})$ and
$x^\sigma_{j} = x^\sigma_{j+1}$ satisfies the
claim.

Otherwise $\sigma(W_{j})\setminus \cR_{j} = \mathset{y}$
for some $y\in [n]$; it would suffice to show
$y=x^\sigma_{j+1}$, since then $\cR_{j}\setminus
\sigma(W_{j}) = \{x^\sigma_{j}\}$ would satisfy
the claim.

Consider then $\sigma_j$, the $C_{j}$-parent of $\sigma$. 
Note that $\sigma_j(W_{j}) = \sigma(W_{j})$, 
and since $C_{j}$ employs \quo{push-to-the-$(k(j-1)+1)$st-index}
transitions only, and
\[\sigma_0(W_{j}\cup\dots\cup W_d) \subseteq
\cR_{j}\cup\dots\cup \cR_d,\]
we know that $\sigma(W_{j})\subseteq \cR_{j}\cup\dots\cup \cR_d$. 
We now use the induction hypothesis
\[\sigma(W_{j+1}\cup\dots\cup W_{d}) = 
\parenv{\cR_{j+1}\cup\dots\cup\cR_d}\setminus\mathset{x^\sigma_{j+1}},\]
and it follows that $\sigma(W_{j})\subseteq \cR_{j} \cup 
\mathset{x^\sigma_{j+1}}$, hence $y=x^\sigma_{j+1}$.
\end{IEEEproof}

From now on, we denote $i^\sigma_j = \sigma^{-1}(x^\sigma_j)$. Another
useful notation we shall employ is a function that quantizes any
integer to the nearest integer leaving a residue of $j$ modulo $d$. We
denote this function by $q_d^j:\Z\to d\Z+j$, defined by
\[ q_d^j(a) = \argmin_{b\in d\Z+j} \abs{ a-b},\]
where we assume $\argmin$ returns a single value, and ties are broken
arbitrarily.

For the decoding procedure description, let us fix the parameters
$n=kd$, and the code $C_1$ from \autoref{con:main}. Additionally, we
denote by $\sigma\in C_1$ the transmitted permutation, by $\tau\in
S_n$ the received permutation, and by $\shat\in S_n$ the decoded
permutation. We denote the \emph{decoding radius} by
$t=\floorenv{(d-1)/2}$, and assume $d_\infty(\sigma,\tau)\leq t$.

We will decode $\tau$ iteratively by window, from $W_1$ to $W_d$. 
We shall make sure--inductively--that when we begin the process of
decoding $W_j$, for some $j\in[d]$, we know $i^\sigma_j$.
Initially, as mentioned, we set $j=1$. Trivially, $i^\sigma_1=1$.

\begin{step}
  \label{dec:step1-naive}
  We set the decoding window
  \[\hat{W}_j = W_j\cup\mathset{i^\sigma_j},\]
  and naively decode $\hat{W}_j$ by setting for all $i\in \hat{W}_j$, 
  \[\shat(i) = q_d^j(\tau(i)).\]
\end{step}

\begin{lemma}
  \label{dec:lem:correct}
  After \autoref{dec:step1-naive}, for all $i\in \hat{W}_j$ such that 
  $\sigma(i)\in\cR_j$ it holds that $\shat(i) = \sigma(i)$. 
\end{lemma}
\begin{IEEEproof}
For all such $i$ we have $\shat(i)\equiv\sigma(i)\pmod d$ and
\begin{align*}
\abs{\shat(i) - \sigma(i)} &\leq \abs{\shat(i) - \tau(i)} + \abs{\tau(i)-\sigma(i)}\\
&= \abs{q_d^j(\tau(i)) - \tau(i)} + \abs{\tau(i)-\sigma(i)}\\
&\leq \floorenv{d/2} + t < d.
\end{align*}
\end{IEEEproof}

\begin{corollary}
  \label{dec:cor:dup}
  After \autoref{dec:step1-naive},
  \[\shat(\hat{W}_j) = 
  \shat\parenv{\hat{W}_j\setminus\mathset{i^\sigma_{j+1}}} = \cR_j.\]
\end{corollary}
\begin{IEEEproof}
By \autoref{cor:missing-element} we know 
that $\cR_j\subseteq \sigma\parenv{\hat{W}_j}$. We further recall that 
$\sigma(i^\sigma_{j+1})\not\in\cR_j$, hence
\[\cR_j\subseteq \sigma\parenv{\hat{W}_j\setminus\mathset{i^\sigma_{j+1}}},\]
and since
\[\abs{\sigma\parenv{\hat{W}_j\setminus\mathset{i^\sigma_{j+1}}}} = 
\abs{\hat{W}_j\setminus\mathset{i^\sigma_{j+1}}} = k = \abs{\cR_j}\]
we have equality. The claim now follows from \autoref{dec:lem:correct}.
\end{IEEEproof}

\autoref{dec:cor:dup} implies that after \autoref{dec:step1-naive}, 
$\shat(\hat{W}_j)$ contains a unique element of $\cR_j$ which appears 
twice, and every other element appears exactly once; by 
\autoref{dec:lem:correct} these other elements have been decoded 
correctly. Before we can continue inductively to decode $W_{j+1}$, 
it only remains to find $i^\sigma_{j+1}$; the other instance 
in $\hat{W}_j$ of $\shat(i^\sigma_{j+1})$ we therefore also know to have been 
decoded correctly.

We shall identify $i^\sigma_{j+1}$ using $\Caux$, the auxiliary
$\Gaux({k+1},\tilde{M}_{k+1})$ code used in \autoref{con:main}. By
construction, if we examine $\sigma_j$, the $C_{j}$-parent of
$\sigma$, then for all $i\in W_j$ we observe $\sigma(i)=\sigma_j(i)$, 
and $\sigma(i^\sigma_j)=\sigma_j(k(j-1)+1)$. The ordering of the $k+1$ 
elements of $\sigma(\hat{W}_j) = \sigma_j\parenv{\mathset{k(j-1)+1}\cup 
W_j}$ is then induced by a permutation of $\Caux$. We
construct this induced permutation from the auxiliary code $\Caux$,
which we denote $\phat\in S_{k+1}$. We first define a simple bijection
$\alpha_j:\cR_j\to [k]$, which is the inverse of the enumeration of
$\cR_j$ given by the arbitrary initial order of elements in $\sigma_0$
used in \autoref{con:main}, e.g., in the simple case $n=kd$, 
\[
\alpha_j(m) = \begin{cases}
  \floorenv{\frac{m}{d}} & j<m\in\cR_j,\\
  k & m=j.
\end{cases}
\]
With $\alpha_j$ we define $\phat$ as,
\[
\phat(i) = \begin{cases}
  \alpha_j(\shat(i^\sigma_j)) & i=1;\\
  \alpha_j(\shat(k(j-1)+i)) & i\in\mathset{2,3,\dots,k+1},
\end{cases}
\]
and note that--as it currently stands--$\phat$ is not a permutation of 
$[k+1]$ because its range is $[k]$ and some unique $a\in [k]$ has 
two distinct pre-images.

\begin{theorem}
  \label{dec:step_naive}
  Let $s,t\in [k+1]$ be the unique pair of indices such that $\phat(s)
  = \phat(t) = a \in [k]$. There is a unique way to re-define
  $\phat\upharpoonright_{\mathset{s,t}}$ (the restriction of $\phat$
  to $\mathset{s,t}$) as a bijection onto $\mathset{a,k+1}$ that
  yields $\phat\in\Caux$. Furthermore, if we define
  $I_j:[k+1]\times[n]\to [n]$ by
  \[I_j(q, r) = \begin{cases}
    r & q=1,\\
    k(j-1) + q & \text{otherwise}
  \end{cases}\]
  then after performing that correction 
  \[i^\sigma_{j+1} = I_j(\phat^{-1}(k+1), i^\sigma_j).\]
\end{theorem}
\begin{IEEEproof}
First, arbitrarily set $\phat(t)=k+1$, where $t>s$. Once corrected, 
$\phat\in S_{k+1}$ by \autoref{dec:cor:dup} and because $\alpha_j:
\cR_j\to [k]$ is a bijection.

Now, we take $\pi\in\Caux$ which generates $\sigma_j$ in the recursion 
step of \autoref{con:main}--while constructing $C_j$--from its 
$C_{j+1}$-parent. Hence 
\[\pi(i) = \begin{cases}
  \alpha_j(x^\sigma_j) & i=1,\\
  \alpha_j(\sigma(k(j-1)+i)) & i\in\mathset{2,3,\dots,k+1},
\end{cases}\]
and therefore either $\phat = \pi$ or $\phat = (k+1, a)\circ\pi$.
Crucially, we observe that in the latter case $\phat\not\in\Caux$
since $\Caux$ is a $\Gaux(k+1,\tilde{M}_{k+1})$ code and $\pi\in
\Caux$; we utilize $\mathtt{ValidAux}$ to discover whether our
original arbitrary correction should be reversed.

To complete the proof, we note by the recursion step of \autoref{con:main} 
that, indeed, $i^\sigma_{j+1} = I_j(\pi^{-1}(k+1), i^\sigma_j)$.
\end{IEEEproof}

We can therefore complete our iterative decoding round with the following 
step.
\begin{step}
  \label{dec:step_cor}
  We construct $\phat$ as described, identify $s,t$, $s<t$, and 
  arbitrarily correct $\phat(t)=k+1$. We test $\mathtt{ValidAux}(\phat)$: 
  if true, we have $i^\sigma_{j+1} = I_j(t, i^\sigma_j)$; otherwise, it 
  holds that $i^\sigma_{j+1} = I_j(s, i^\sigma_j)$.
\end{step}

Finally, observe that when decoding $W_d$ it's known that
$\sigma(\hat{W}_d) = \cR_d$, hence by \autoref{dec:lem:correct} 
$\hat{W}_d$ is decoded correctly, and we need not (and--indeed--cannot) 
perform \autoref{dec:step_cor}.

\begin{example}
We shall demonstrate the decoding process assuming once again $n=kd$
for simplicity, and using the parameters $d=3$ (hence $t=1$), $k=2$
and code constructed in \autoref{exm:con:rec}. Recall that the
$\Gaux(3,3)$ code used in that example is
\[\Caux = \mathset{\sparenv{1,2,3},\ \sparenv{3,1,2},\ \sparenv{2,3,1}}.\]
We choose the transmitted codeword $\sigma = \sparenv{1,2,4,6,5,3}$, and 
a noisy received permutation $\tau = \sparenv{1,3,4,5,6,2}$.

We start by defining $i_1 = 1$ and observing (by abuse of the vector
notation) $\tau\upharpoonright_{\hat{W}_1} = \sparenv{1;3,4}$ (the
first element is differentiated because--generally although never when
$j=1$--it does not immediately precede the rest in $\tau$'s vector
notation).

Since $j=1$, we define $\shat\upharpoonright_{\hat{W}_1} = \sparenv{1;4,4}$. 
This leads us to construct $\phat = \sparenv{2,1,3}\not\in\Caux$, so we 
instead correct $\phat = \sparenv{2,3,1}$ and define $i_2 = 2$. (So far 
we have $\shat = \sparenv{1,\underline{4},4,\cdot,\cdot,\cdot}$, where 
an underline marks $i^\sigma_{j+1}$.)

Next, we have $\tau\upharpoonright_{\hat{W}_2} = \sparenv{3;5,6}$, which 
($j=2$) we decode $\shat\upharpoonright_{\hat{W}_1} = \sparenv{2;5,5}$. 
This again generates $\phat = \sparenv{2,1,3}\not\in\Caux$, and we correct 
in similar fashion to $\phat = \sparenv{2,3,1}$ and define $i_2 = 4$.
(Up to this point, we have $\shat = \sparenv{1,\cancel{4}2,4,\underline{5},
5,\cdot}$).

Finally, we have $\tau\upharpoonright_{\hat{W}_3} = \sparenv{5;2}$ 
and since $j=3$ we decode $\shat\upharpoonright_{\hat{W}_1} = 
\sparenv{6;3}$, and overall $\shat = \sparenv{1,2,4,\cancel{5}6,5,3} = 
\sigma$.
\end{example}

\begin{example}
We present another example, intended to demonstrate the process in
more detail, for which we depart from the parameters used in
\autoref{exm:con:rec} by setting $d=5$ (allowing for
$t=2\leq\floorenv{(d-1)/2}$), $k=3$. In each recursion step of
\autoref{con:main} the $\Gaux(4, 8)$ code used is $\Caux$ presented in
\autoref{exm:aux}.

The codeword
\[\sigma = \sparenv{11,1,8,6,7,2,12,13,3,5,9,14,4,10,15}\]
appears in the code generated in this case, as can be seen by identifying 
its $C_5$, $C_4$, $C_3$, and $C_2$ parents as, respectively,
\begin{align*}
\sigma_5 &= \sparenv{6,11,1,7,12,2,8,13,3,9,14,4,5,10,15},\\
\sigma_4 &= \sparenv{6,11,1,7,12,2,8,13,3,5,9,14,4,10,15},\\
\sigma_3 &= \sigma_4,\\
\sigma_2 &= \sparenv{6,11,1,8,7,2,12,13,3,5,9,14,4,10,15}.
\end{align*}
We choose
\[\tau = \sparenv{12,3,9,7,5,2,11,15,1,6,8,13,4,10,14}\]
to be the noisy version of the transmitted codeword $\sigma$, and
verify that $d_\infty(\tau,\sigma) = 2 = t$.

Beginning with $j=1$, we have $\tau\upharpoonright_{\hat{W}_1} = 
\sparenv{12;3,9,7}$, which we decode $\shat\upharpoonright_{\hat{W}_1} = 
\sparenv{11;1,11,6}$, generating $\phat = \sparenv{2,3,2,1}$ which is 
corrected to $\phat = \sparenv{2,3,4,1}\in\Caux$. We identify $i_2 = 3$, 
and keep
\[\shat = \sparenv{11,1,\underline{11},6,\cdot,\cdot,\cdot,\cdot,
\cdot,\cdot,\cdot,\cdot,\cdot,\cdot,\cdot}.\]

Next, for $j=2$, observe that $\tau\upharpoonright_{\hat{W}_2} =
\sparenv{9;5,2,11}$, and we decode $\shat\upharpoonright_{\hat{W}_2} =
\sparenv{7;7,2,12}$.  This generates $\phat = \sparenv{1,1,3,2}$,
which we initially correct to $\phat = \sparenv{1,4,3,2}\not\in\Caux$,
so (skip correcting $\phat$, as it has no further consequence) $i_3 =
i_2 = 3$ instead of $i_3 = 4$.
We summarize
\[\shat =
\sparenv{11,1,\cancel{11}\underline{7},6,7,2,12,
  \cdot,\cdot,\cdot,\cdot,\cdot,\cdot,\cdot,\cdot}.\]

We turn to $\hat{W}_3$ and see that $\tau\upharpoonright_{\hat{W}_3} = 
\sparenv{9;15,1,6}$, decoded to $\shat\upharpoonright_{\hat{W}_3} = 
\sparenv{8;13,3,8}$. We generate $\phat = \sparenv{1,2,3,1}$ and correct 
it to $\phat = \sparenv{1,2,3,4}\in\Caux$, indicating that $i_4 = 10$. 
We now have
\[\shat = \sparenv{11,1,\cancel{7}8,6,7,2,12,13,3,
\underline{8},\cdot,\cdot,\cdot,\cdot,\cdot}.\]

Moving on to $j=4$, while decoding $\hat{W}_4$ we note
$\tau\upharpoonright_{\hat{W}_4} = \sparenv{6;8,13,4}$, which we
decode as $\shat\upharpoonright_{\hat{W}_4} =
\sparenv{4;9,14,4}$. This generates $\phat = \sparenv{3,1,2,3}$ which
is corrected to $\phat = \sparenv{3,1,2,4}\not\in\Caux$. We therefore
define $i_5 = i_4 = 10$ instead of $i_5 = 13$. Up to now,
\[\shat =
\sparenv{11,1,
  8,6,7,2,12,13,3,\cancel{8}\underline{4},9,14,4,\cdot,\cdot}.\]

Finally, $j=5$, and we get $\tau\upharpoonright_{\hat{W}_5} =
\sparenv{4;10,14}$ which we decode to
$\shat\upharpoonright_{\hat{W}_5} = \sparenv{5;10,15}$, and overall
\[\shat = \sparenv{11,1,8,6,7,2,12,13,3,\cancel{4}5,9,14,4,10,15} =
\sigma.\]
\end{example}

\begin{function}[t]
\begin{footnotesize}
\DontPrintSemicolon
\Input{$\tau\in S_{kd}$ satisfying $d_\infty(\tau, C_1)\leq t\leq \floorenv{(d-1)/2}$.}
\Output{$\shat\in C_1$ such that $d_{\infty}(\tau,\shat)\leq t$.}

	$i\gets 1$\\
	\For {$j = 1,2,\ldots,d-1$}
	{
		\tcc{Naively decode $\hat{W}_j$}
		$\shat(i)\gets q_d^j(\tau(i))$\\
		$\phat(1)\gets \alpha_j(\shat(i))$\\
		\For {$r = 2,\ldots, k+1$}
		{
			$m\gets q_d^j(\tau(k(j-1)+r))$\\
			\If {$\shat^{-1}(m)$ is already set}
			{
				$\phat(r)\gets k+1$\\
				$a\gets \alpha_j(m)$\\
			}
			\Else
			{
				$\phat(r)\gets \alpha_j(m)$\\
			}
			$\shat(r)\gets m$\\
		}
		
		\tcc{Define $i_{j+1}$}
		\If {$\mathtt{ValidAux}(\phat)$}
		{
			$i\gets I(\phat^{-1}(k+1), i)$\\
		}
		\Else
		{
			$i\gets I(\phat^{-1}(a), i)$\\
		}
	}
	
	\tcc{Decode $\hat{W}_d$}
	$\shat(i)\gets q_d^d(\tau(i))$\\
	\For {$r = 2,\ldots, k$}
	{
		$\shat(r)\gets q_d^j(\tau(k(d-1)+r))$\\
	}
	
	\Return $\shat$ \nllabel{alg:decode:return}\\
\end{footnotesize}
\caption{() $\mathtt{Decode}\parenv{\tau}$\label{alg:decode}}
\end{function}

The decoding algorithm is formalized in $\mathtt{Decode}(\tau)$. With
appropriate simple data structures, the algorithm requires
$O(kd)=O(n)$ steps. We assume simple integer operations to take
constant-time.

\section{Ranking and Unranking}
\label{sec:ranking}

In this section we discuss the process of encoding data
$m\in\mathset{0,1,\ldots,\abs{C_1}-1}$ to a codeword $\sigma\in C_1$,
which is also known as \emph{unranking} $m$, and the inverse process
of \emph{ranking} $\sigma\in C_1$, i.e., obtaining its rank in the
code. Throughout this section, $C_1$ stands for the code obtained via
\autoref{con:main}.

Due to the nature of our construction, performing these tasks with 
the codes generated by \autoref{iter_const} is reliant on our ability 
to do the same with the codes provided by \autoref{jsnakes} and 
\autoref{auxcodes}. We therefore recall the following known result.

\begin{lemma}\label{jsnakes:rur}\cite{JiaMatSchBru09}
The complete $G_\uparrow(n,n!)$ codes provided by \autoref{jsnakes} 
has a ranking algorithm operating in $O(n)$ steps, and an unranking 
scheme operating in $O(n^2)$ steps.
\end{lemma}

This gives rise to the following corollary.

\begin{corollary}
  The $\Gaux(2m,\frac{\abs{S_{2m}}}{2m-1})$ codes generated 
  by \autoref{flipconst} can be ranked in $O(m)$ operations and 
  unranked in $O(m^2)$ operations.
\end{corollary}
\begin{IEEEproof}
Ranking a permutation $\sigma$ in the code may proceed by finding
the cyclic shift required for $[2m,1]$ to be the first two
elements.  After removing these two first elements, and then
reversing the permutation we may use a ranking algorithm from
\autoref{jsnakes:rur}. A simple combination of the results gives the
required ranking of $\sigma$. By \autoref{jsnakes:rur}, the entire
procedure takes $O(m)$ operations. A symmetric argument gives an
$O(m^2)$ algorithm for unranking.
\end{IEEEproof}
  
Unfortunately, no ranking and unranking schemes are known for 
parity-preserving $G_\uparrow(2m+1, M_{2m+1})$ codes provided by
\autoref{holsnakes} (developed in \cite{Hol16}), or previous
constructions presented in \cite{HorEtz14,ZhaGe16}. Consequentially, 
we rely on \autoref{flipconst} instead of \autoref{aux:final} for 
even sized congruence classes. In the case of odd sized classes, we 
can leverage the following codes.

\begin{lemma}\label{ysnakes}\cite{YehSch12b}
For all $m\geq 1$ there exist parity-preserving
${G_\uparrow(2m+1,\hat{M}_{2m+1})}$ codes with sizes
\[\hat{M}_{2m+1} = \parenv{\frac{(2m)!}{m!}}^2\frac{(2m+1)}{2^{2m}} = \frac{(2m)!}{m!^2 2^{2m}}\abs{S_{2m+1}}.\]
These codes can be ranked and unranked in $O(m^2)$ operations.
\end{lemma}

We summarize those observations in the following corollary.

\begin{corollary}\label{auxcodes:rur}
For all $k\geq 3$ there exist a $\Gaux(k,\hat{M}_k)$ code 
starting with $\id$ and a $t_{\uparrow k}$ transition, which have 
ranking and unranking schemes operating in $O(k^2)$ steps, where 
\[\hat{M}_k = \begin{cases}
\frac{(k-1)!}{((k-1)/2)!^2 2^{k-1}}k! & k\equiv 1\pmod{2};\\
\frac{k!}{(k-1)} & k\equiv 0\pmod{2}.
\end{cases}\]
\end{corollary}

Note that we can now replace \autoref{auxcodes} by
\autoref{auxcodes:rur} in \autoref{con:main} to obtain codes which we
shall denote $\hat{C}_1$, and each auxiliary code on a congruence 
class of size $k>1$ contributes to $\abs{\hat{C}_1}$ a multiplicative 
factor of
\[\hat{M}_{k+1} = \begin{cases}
\frac{k!}{(k/2)!^2 2^k}(k+1)! & k\equiv 0\pmod 2,\\
\frac{(k+1)!}{k} & k\equiv 1\pmod 2.
\end{cases}\]

We also note, using Stirling's approximation
\[e^{\frac{1}{12n+1}} < \frac{n!e^n}{n^n\sqrt{2\pi n}} < e^{\frac{1}{12n}},\]
that
\[
\frac{k!}{(k/2)!^2 2^k} > 
\sqrt{\frac{2}{\pi k}} \parenv{e^{1-\frac{1}{4 + \frac{1}{3k}}}}^{-\frac{1}{3k}} >
\sqrt{\frac{2}{\pi k}} e^{-1/(4k)}.
\]

We can then recalculate code size, in the case of 
$\floorenv{\frac{n}{d}}\equiv 0\pmod 2$:
\begin{align*}
\abs{\hat{C}_1} &= 
\parenv{\frac{(\ceilenv{n/d}+1)!}{\ceilenv{n/d}}}^{n\bmod d}\cdot
\floorenv{\frac{n}{d}}!\\
&\quad\ \cdot\parenv{
\frac{\floorenv{n/d}!}{(\floorenv{n/d}/2)!^2 2^{\floorenv{n/d}}}
\parenv{\floorenv{\frac{n}{d}}+1}!}^{d-(n\bmod d)-1}\\
&> \left.\ceilenv{\frac{n}{d}}!\right.^{n\bmod d}
\left.\floorenv{\frac{n}{d}}!\right.^{d-\parenv{n\bmod d}}\\
&\quad\  \cdot 
\parenv{1+\frac{1}{\ceilenv{n/d}}}^{n\bmod d}\cdot 
\parenv{\floorenv{\frac{n}{d}}+1}^{(d-1)-(n\bmod d)}\\
&\quad\ \cdot\parenv{\sqrt{\frac{2}{\pi \floorenv{n/d}}} e^{-1/(4\floorenv{n/d})}}^{(d-1)-(n\bmod d)},
\end{align*}
and when $\floorenv{\frac{n}{d}}\equiv 1\pmod 2$:
\begin{align*}
\abs{\hat{C}_1} &= 
\parenv{
\frac{\ceilenv{n/d}!}{(\ceilenv{n/d}/2)!^2 2^{\ceilenv{n/d}}}
\parenv{\ceilenv{\frac{n}{d}}+1}!}^{n\bmod d}\\
&\quad\  \cdot \floorenv{\frac{n}{d}}!\cdot
\parenv{\frac{(\floorenv{n/d}+1)!}{\floorenv{n/d}}}^{d-(n\bmod d)-1}\\
&> \left.\ceilenv{\frac{n}{d}}!\right.^{n\bmod d}
\left.\floorenv{\frac{n}{d}}!\right.^{d-\parenv{n\bmod d}}\\
&\quad\  \cdot 
\parenv{\ceilenv{\frac{n}{d}}+1}^{n\bmod d}\cdot 
\parenv{1+\frac{1}{\floorenv{n/d}}}^{(d-1)-(n\bmod d)}\\
&\quad\  \cdot\parenv{\sqrt{\frac{2}{\pi \ceilenv{n/d}}} e^{-1/(4\ceilenv{n/d})}}^{n\bmod d},
\end{align*}
and we note that in the special case $\floorenv{\frac{n}{d}}=1$ 
we have $\abs{\hat{C}_1}=\abs{C_1}$.

We likewise observe the rates of codes based on \autoref{auxcodes:rur}, 
and find for $\floorenv{1/\delta}\equiv 0\pmod{2}$
\begin{align*}
\hat{R} &\geq \parenv{1 - \delta\floorenv{\frac{1}{\delta}}}
\log_2\parenv{\ceilenv{\frac{1}{\delta}}!\parenv{1+\frac{1}{\ceilenv{1/\delta}}}}\\
&\quad\  + \parenv{\delta + \delta\floorenv{\frac{1}{\delta}} - 1}
\log_2\parenv{\parenv{\floorenv{\frac{1}{\delta}}+1}!}\\
&\quad\ - \frac{1}{2}\parenv{\delta + \delta\floorenv{\frac{1}{\delta}} - 1}\\
&\quad\  \cdot \parenv{\log_2\floorenv{\frac{1}{\delta}} + \frac{\log_2(e)}{2\floorenv{1/\delta}} + \log_2(\pi) - 1} - o(1),
\end{align*}
and for $\floorenv{1/\delta}\equiv 1\pmod{2}$
\begin{align*}
\hat{R} &\geq \parenv{1 - \delta\floorenv{\frac{1}{\delta}}}
\log_2\parenv{\parenv{\ceilenv{\frac{1}{\delta}}+1}!}\\
&\quad\  + \parenv{\delta + \delta\floorenv{\frac{1}{\delta}} - 1}
\log_2\parenv{\floorenv{\frac{1}{\delta}}!\parenv{1+\frac{1}{\floorenv{1/\delta}}}}\\
&\quad\  - \frac{1}{2}\parenv{1 - \delta\floorenv{\frac{1}{\delta}}}\\
&\quad\  \cdot \parenv{\log_2\ceilenv{\frac{1}{\delta}} + \frac{\log_2(e)}{2\ceilenv{1/\delta}} + \log_2(\pi) - 1} - o(1).
\end{align*}

\begin{figure*}[ht]
\psfrag{xbx}{\small{(a)}}
\psfrag{xcx}{\small{(b)}}
\psfrag{xex}{\small{(c)}}
\psfrag{rrr}{$R$}
\psfrag{delta}{$\delta$}
\vspace*{-20ex}
\includegraphics[width=\linewidth]{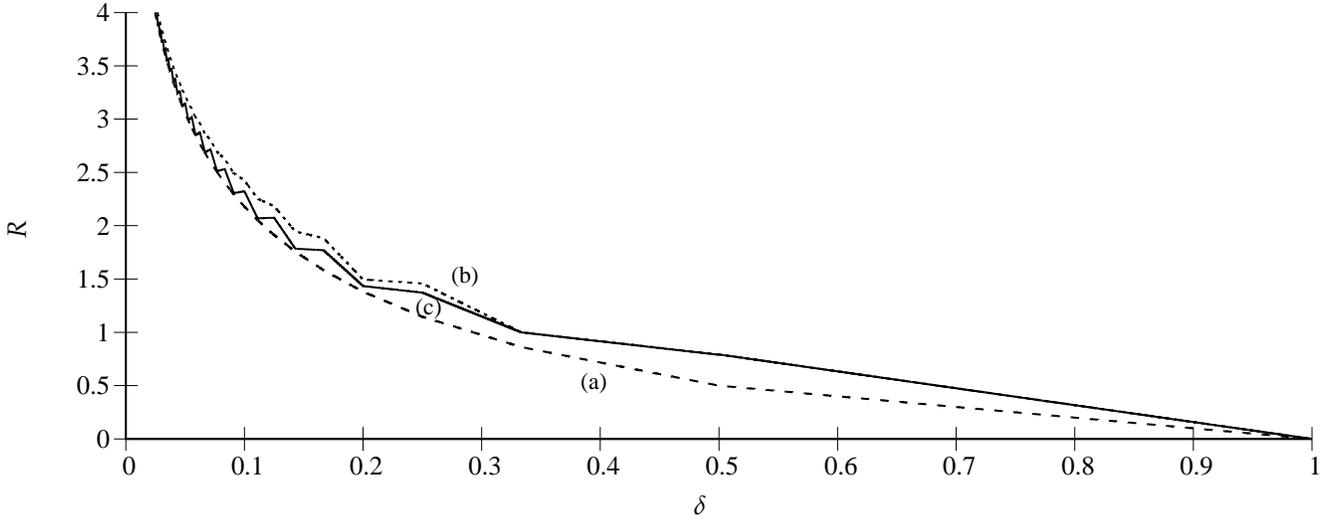}
\vspace*{-20ex}
\caption{(a) The rate of codes from \autoref{const:Tam} constructed in 
  \cite{TamSch10}. (b) The rate of codes $C_1$ from
  \autoref{con:main}. (c) The rate of codes $\hat{C}_1$ constructed 
  using auxiliary codes from \autoref{auxcodes:rur}.}
\label{figrankable}
\end{figure*}

The losses in asymptotic rate are shown in \autoref{figrankable}. We
observe in particular that we still manage to achieve better rates
than previously known error-correcting codes (without the Gray
property), even with the significantly smaller
$\Gaux(k,\hat{M}_k)$ of \autoref{auxcodes:rur}.

Let us denote by $\mathtt{RankComplete}(\pi)$,
$\mathtt{UnrankComplete}(m)$ the ranking and unranking procedures for
the complete codes from \autoref{jsnakes:rur}. Additionally, let
$\mathtt{RankAux}(\pi)$ and $\mathtt{UnrankAux}(m)$ denote the ranking
and unranking procedures for the auxiliary codes of
\autoref{auxcodes:rur}. We can readily take advantage of
$\hat{C}_1$'s tiered structure to use these functions in order to
perform the same tasks for our construction. We include pseudo-code
for these algorithms, which we call $\mathtt{Rank}(\sigma)$ and
$\mathtt{Unrank}(m)$, for completeness. As before, we assume $n=kd$ to
simplify the presentation.

\begin{function}[t]
\begin{footnotesize}
\DontPrintSemicolon
\Input{$\sigma\in \hat{C}_1$.}
\Output{$m\in\mathset{0,1,\ldots,\abs{\hat{C}_1}-1}$ which is the rank of $\sigma$ in $\hat{C}_1$.}

	\tcc{Build a permutation $\pi_d\in S_k$}
	\For {$i\in W_d$}
	{
		$\pi_d[i-k(d-1)]\gets \alpha_d(\sigma[i])$ \\
	}
	$\pi_d[1]\gets [k]\setminus\pi_d[2,\ldots,k]$\\
	$m \gets \parenv{(\mathtt{RankComplete}(\pi_d)-1) \bmod k!}$\\
	
	\For {$j=d-1,\ldots, 1$}
	{
		\tcc{Build a permutation $\pi_j\in S_{k+1}$}
		\For {$i\in W_j$}
		{
			$\pi_j[i-k(j-1)]\gets \alpha_j(\sigma[i])$\\
		}
		$\pi_j[1]\gets [k+1]\setminus\pi_j[2,\ldots,k+1]$\\
		$m \gets m\cdot\hat{M}_{k+1}+\parenv{(\mathtt{RankAux}(\pi_j)-1) \bmod \hat{M}_{k+1}}$\\
	}
	
	\Return $(m+1)\bmod \abs{\hat{C}_1}$ \nllabel{alg:rank:return}\\
\end{footnotesize}
\caption{() $\mathtt{Rank}\parenv{\sigma}$\label{alg:rank}}
\end{function}

\begin{function}[t]
\begin{footnotesize}
\DontPrintSemicolon
\Input{$m\in\mathset{0,1,\ldots,\abs{\hat{C}_1}-1}$.}
\Output{$\sigma\in \hat{C}_1$ with rank $m$ in $\hat{C}_1$.}

	\tcc{Convert $m$ to local ranks $\mathtt{R}[1,2,\dots,d]$}
	$m\gets \parenv{(m-1) \bmod \abs{\hat{C}_1}}$\\
	\For {$i=1,2,\dots,d-1$}
	{
		$\mathtt{R}[i]\gets \parenv{(m+1) \bmod \hat{M}_{k+1}}$\\
		$m\gets \floorenv{ m / \hat{M}_{k+1} }$\\
	}
	$\mathtt{R}[d]\gets \parenv{(m+1) \bmod k!}$\\

	\tcc{Construct $\sigma$}
	$\pi_d\gets \mathtt{UnrankComplete}(\mathtt{R}[d])$\\
	\For {$i\in W_d$}
	{
		$\sigma[i]\gets \pi_d[i-k(d-1)]\cdot d$\\
	}
	$x\gets \pi_d[1]\cdot d$\\
	
	\For {$j=d-1,\ldots, 1$}
	{
		$\pi_j\gets \mathtt{UnrankAux}(\mathtt{R}[j])$\\
		\For {$i\in W_j$}
		{
			\If {$\pi_j[i]=k+1$}
			{
				$\sigma[i]\gets x$\\
			}
			\Else
			{
				$\sigma[i]\gets \pi_j[i-k(j-1)]\cdot d + j$\\
			}
		}
		\If {$\pi_j[1]\neq k+1$}
		{
			$x\gets \pi_j[1]\cdot d + j$\\
		}
	}
	$\sigma[1]\gets x$\\
	
	\Return $\sigma$ \nllabel{alg:unrank:return}\\
\end{footnotesize}
\caption{() $\mathtt{Unrank}\parenv{m}$\label{alg:unrank}}
\end{function}

\begin{theorem}\label{rur:complexity}
  For the code $\hat{C}_1$ of length $n=kd$, the algorithms
  $\mathtt{Rank}(\sigma)$, $\mathtt{Unrank}(m)$ operate in $O(k^2 d)$
  steps.
\end{theorem}
\begin{IEEEproof}
Both algorithms perform a single loop over all indices of $\sigma$, 
making simple integer operations, which requires $O(n)$ steps. They
also make a call to one of $\mathtt{RankComplete}(\pi)$, 
$\mathtt{UnrankComplete}(m)$ and $(d-1)$ calls to one of 
$\mathtt{RankAux}(\pi)$, $\mathtt{UnrankAux}(m)$, costing 
$O(k^2)$ operations each.
\end{IEEEproof}

We also note in particular that in the regime $d=\Theta(n)$, we have
$k=\Theta(1)$, and \autoref{rur:complexity} yields linear run-time
$O(n)$.

\section{Snake-in-the-box codes in $S_{2m+2}$}\label{sec:app}

As mentioned before in \autoref{sec:auxcodes}, the issue of asymmetry 
between \quo{push-to-the-top} codes in the symmetric group of odd and 
even orders has also frustrated research into error-detecting codes 
under the Kendall $\tau$-metric in the past.

The \emph{Kendall $\tau$-metric} \cite{KenGib90} on $S_n$ is defined as
\[d_{\cK}(\sigma,\tau) = 
\abs{\mathsetp{(i,j)}{\sigma(i)<\sigma(j)\wedge\tau(i)>\tau(j)}}.\]
Informally, as noted in \cite{JiaSchBru10}, it measures the minimal 
number of adjacent transpositions required to transform one permutation 
into the other, that is, the minimal $r$ such that
\[\sigma = \tau\circ(i_1,i_1+1)\circ(i_2,i_2+1)\circ\ldots,
\circ(i_r,i_r+1)\]
for some $i_1,i_2,\ldots,i_r\in[n-1]$. An $(n,M,\cK)$-snake, or 
{$\cK$-snake} for short, is a single-error-detecting rank-modulation 
Gray code of size $M$, or more formally, a $G_\uparrow(n,M)$ code $C$ 
such that for all $\sigma,\tau\in C$, $\sigma\neq\tau$, it holds that 
$d_{\cK}(\sigma,\tau) \geq 2$. Put differently, for no $i\in[n-1]$ 
does it hold that $\sigma = \tau\circ(i,i+1)$.

The authors have shown in \cite{YehSch12b}[Thm.~17] that any $\cK$-snake 
$C\subseteq S_n$ which employs a \quo{push-to-the-top} transition on an even 
index $t_{\uparrow 2m}$--for any $m\in\floorenv{\frac{n}{2}}$--must satisfy 
$\abs{C}\leq\frac{n!}{2}-\Theta(n)$. 
Horovitz and Etzion posited in \cite{HorEtz14} that $\cK$-snakes in 
$S_{2m+2}$ do not exceed the size of those in $S_{2m+1}$, a conjecture 
refuted when Zhang and Ge demonstrated in \cite{ZhaGe17} the existence 
of $\cK$-snakes in $S_{2m+2}$ of size $\frac{(2m+2)!}{4}$. Concurrently 
and independently, Holroyd conjectured in \cite{Hol16} that $\cK$-snakes 
can be found in $S_{2m+2}$ with size greater than 
$\frac{(2m+2)!}{2}-O(m^2)$.

A resemblance is evident in the definitions of $(n,M,\cK)$-snakes and 
$\Gaux(n,M)$ codes, which is reinforce by the observations that, 
similarly to properties seen in \autoref{sec:auxcodes}, any 
parity-preserving $G_\uparrow(n,M)$ code is an $(n,M,\cK)$-snake (see 
\cite{YehSch12b}[Lem.~5]), and any $(n,M,\cK)$-snake satisfies 
$M\leq\frac{n!}{2}$ (see \cite{YehSch12b}[Thm.~15]).

We wish to demonstrate how the principles behind \autoref{aux:final} can 
be applied to the construction of a $\cK$-snake in $S_{2m+2}$ of size 
$M\approx\frac{(2m+2)!}{2}$.

\begin{lemma}\label{hsnakes}\cite[Thm.~18]{HorEtz14}\cite{ZhaGe16}
For $m\geq 2$, there exist parity-preserving 
$G_\uparrow(2m+1,M_{2m+1})$ codes with
\[M_{2m+1} = \abs{A_{2m+1}}-(2m-1) = \frac{(2m+1)!}{2}-(2m-1).\]
In particular, such a code $C$ was constructed such that, as a group, 
\[C = A_{2m+1} \setminus \mathset{t_{\uparrow 2m-1}{}^q\sigma}_{q=0}^{2m-2}\]
for some $\sigma\in A_{2m+1}$.
Finally, $C$ only employed $t_{\uparrow 2m-1}$, $t_{\uparrow 2m+1}$.
\end{lemma}

As before, we fix $m\geq 2$. We also reuse
\begin{align*}
\varphi(\pi) &= t_{\uparrow 2m+2}{}^2\circ t_{\uparrow 2m-1}{}^{-1}(\pi)\\
&= \pi\circ (1,2m+1)(2m+2,2m,2m-1,\ldots,2)
\end{align*}
and the permutations $\phat_r = \varphi^r(\id)$.

\begin{theorem}
  \label{sna:cycle}
  For all $r\geq 0$ a parity-preserving 
  $G_\uparrow\parenv{2m+2,\frac{(2m+1)!}{2}-(2m-1)}$ code $\hat{P}_r$ 
  exists which satisfy:
  \begin{enumerate}
  \item
  The first permutation in $\hat{P}_r$ is $\phat_r$.
  
  \item
  The last permutation in $\hat{P}_r$ is $t_{\uparrow 2m-1}{}^{-1}\phat_r$.
  
  \item
  For all $\pi\in\hat{P}_r$ it holds that 
  \begin{align*}
  \pi&(2m+2) = \phat_r(2m+2)\\
  &= \begin{cases}
  2m+2 & r\equiv 0\pmod{2m},\\
  2m+1 - \parenv{r\bmod 2m} & r\not\equiv 0\pmod{2m}.
  \end{cases}
  \end{align*}
  
  \item
  $\tilde{\sigma}_r \not\in \hat{P}_r$, where we denote
  \[\tilde{\sigma}_r = \parenv{t_{\uparrow 2m+2}{}^{-1}\phat_r}
  \circ (2m+1,2m+2)\]
  (and observe $\tilde{\sigma}_r = t_{\uparrow 2m+1}{}^{-1}(\phat_r)$, 
  hence in particular $\tilde{\sigma}_r(2m+2) = \phat_r(2m+2)$).
  \end{enumerate}
\end{theorem}
\begin{IEEEproof}
By \autoref{hsnakes} we know that there exist a parity-preserving 
$G_\uparrow(2m+1,M_{2m+1})$ code $P$ such that, as a group, 
\[P = S_{2m+1} \setminus \mathset{t_{\uparrow 2m-1}{}^q\sigma}_{q=0}^{2m-2}\]
for some $\sigma\in A_{2m+1}$.
% satisfying $\sigma(2m)=1$, $\sigma(2m+1)=2$.
We also know that $P$ only employs $t_{\uparrow 2m-1}$, 
$t_{\uparrow 2m+1}$ transitions.

We apply its generating sequence to $\phat_r$ to generate the 
$G_\uparrow(2m+2,M_{2m+1})$ code $\hat{P}$, which employs only 
$t_{\uparrow 2m-1}$, $t_{\uparrow 2m+1}$ transitions (in particular, 
it never employs $t_{\uparrow 2m+2}$, hence  point 3 is established), 
and note that as a group 
\begin{align*}
\hat{P} =& \mathsetp{\tau\in A_{2m+2}}{\tau(2m+2) = \phat_r(2m+2)}\\
&\setminus \mathset{t_{\uparrow 2m-1}{}^q\shat}_{q=0}^{2m-2}
\end{align*}
for some $\shat\in A_{2m+2}$, satisfying 
$\shat(2m+2)=\phat_r(2m+2)$.

Denote $\hat{P} = \parenv{c_j}_{j=1}^{M_{2m+1}}$. We modify our code by 
defining  
\[\hat{P}_r = \parenv{c^\prime_j}_{j=1}^{M_{2m+1}} = 
\parenv{\tilde{\sigma}_r\shat^{-1}c_j}_{j=1}^{M_{2m+1}},\]
which is still a $G_\uparrow(2m+2,M_{2m+1})$ since \quo{push-to-the-top} 
transitions are group-actions by right-multiplication. Moreover, since 
$\tilde{\sigma}_r(2m+2) = \shat(2m+2) = \phat_r(2m+2)$, as a group we 
have 
\begin{align*}
\hat{P}_r =& \mathsetp{\tau\in A_{2m+2}}{\tau(2m+2) = \phat_r(2m+2)}\\
&\setminus \mathset{t_{\uparrow 2m-1}{}^q\tilde{\sigma}_r}_{q=0}^{2m-2}.
\end{align*}
Note in particular that
\[\tilde{\sigma}_r(2m+1) = \phat_r(1)\neq \phat_r(2m+1),\]
hence $\phat_r\in\hat{P}_r$. In addition, point 4 is thus substantiated.

Finally, $t_{\uparrow 2m+1}{}^{-1}(\phat_r) = \tilde{\sigma}_r\not\in 
\hat{P}_r$ implies that $\phat_r$ must necessarily be preceded in 
$\hat{P}_r$ by $t_{\uparrow 2m-1}$, which substantiates point 2 
(after a proper cyclic shift of $\hat{P}_r$).
\end{IEEEproof}

As in \autoref{sec:auxcodes}, $\hat{P}_r\subseteq A_{2m+2}$ for all $r$. 
We construct a $(2m+2,M,\cK)$-snake by stitching together $\hat{P}_0, 
\hat{P}_1, \ldots, \hat{P}_{2m-1}$ in the following lemma.

\begin{lemma}
  \label{sna:stitch}
  For all $r\geq 0$, we may concatenate $\hat{P}_r, \hat{P}_{r+1}$ into 
  a (non-cyclic) \quo{push-to-the-top} code by applying the 
  transitions $t_{\uparrow 2m+2}, t_{\uparrow 2m+2}$ to the last 
  permutation of $\hat{P}_r$, which is $t_{\uparrow 2m-1}{}^{-1}\phat_r$.
  
  The only odd permutation in the resulting code is then 
  \[\beta_{r+1} = t_{\uparrow 2m+2}{}^{-1}(\phat_{r+1}),\]
  which we again call the $(r+1)$-bridge.
\end{lemma}
\begin{IEEEproof}
Exactly as in the proof of \autoref{aux:stitch}, given that $P_r$, 
$\hat{P}_r$ are parity-preserving, and have the same first and last 
permutations.
\end{IEEEproof}

Again, similarly to \autoref{sec:auxcodes}, \autoref{sna:stitch} can be 
used iteratively to cyclically concatenate $\hat{P}_0, \hat{P}_1, \ldots, 
\hat{P}_{2m-1}$, with a single odd permutation--the $r$-bridge--between 
$\hat{P}_{(r-1)\bmod 2m}, \hat{P}_{r}$. Let us prove that fact in the 
following theorem.

\begin{theorem}
  \label{sna:final}
  There exists a $(2m+2,\check{M}_{2m+2},\cK)$-snake for all 
  $m\geq 2$, with 
  \[\check{M}_{2m+2} = \frac{2m}{2m+2}\cdot\frac{(2m+2)!}{2} - (2m-2)2m\]
\end{theorem}
\begin{IEEEproof}
We define $P$, similarly to \autoref{sec:auxcodes}, as the cyclic 
concatenation
\[\hat{P}_0, \beta_1, \hat{P}_r, \beta_2, \dots, \beta_{2m-1}, 
\hat{P}_{2m-1}, \beta_0.\]

Suppose $\pi_1,\pi_2\in C$ satisfy
\[\pi_1 = \pi_2\circ(i,i+1)\]
for some $i\in[2m+1]$, then w.l.o.g $\pi_2$ is odd and hence 
$\pi_2 = \beta_r$ for some $0\leq r < 2m$, and $\pi_1$ is even 
and thus not a bridge; it must follow, then, that 
\[\pi_2(2m+2)\in\mathset{1,2m+1}\not\ni \pi_1(2m+2),\]
hence $i=2m+1$ and 
\begin{align*}
\pi_1 &= \pi_2\circ (2m+1,2m+2)\\
&= \parenv{t_{\uparrow 2m+2}{}^{-1}(\phat_r)}\circ (2m+1,2m+2)\\
&= t_{\uparrow 2m+1}{}^{-1}(\phat_r) = \tilde{\sigma}_r.
\end{align*}
This is in contradiction to \autoref{sna:cycle}, since $\pi_1(2m+2) = 
\phat_r(2m+2)$ and thus $\pi_1\in\hat{P}_r$. Hence $\hat{P}$ is a 
$\cK$-snake. Now, that
\begin{align*}
\abs{\hat{P}} &= 2m\sparenv{\frac{(2m+1)!}{2} - (2m-1)} + 2m\\
&= \frac{2m}{2m+2}\cdot\frac{(2m+2)!}{2} - (2m-2)2m
\end{align*}
is trivial.
\end{IEEEproof}

To conclude this section, we note that 
$\frac{\check{M}_{2m+2}}{\abs{S_{2m+2}}}\tends{m\to\infty} \frac{1}{2}$, 
which is optimal. The authors are unaware of any current result achieving 
this. We add that, in particular, in the context of $\cK$-snakes it is 
common to define the \emph{rate} of codes as $R = \lim_{m\to\infty} 
\frac{\log \abs{\check{M}_{2m+2}}}{\log\abs{S_{2m+2}}}$ (see 
\cite{YehSch12b}), and we naturally observe that in our case $R=1$ 
(which, again, is optimal, although $R=1$ is also achieved by existing 
constructions, e.g., that of \cite{ZhaGe17}).

\section{Conclusion}\label{sec:conclusion}

In this paper we presented the class of $\Gaux(k,M)$ codes,
leveraging codes designed for the rank-modulation scheme under
the Kendall $\tau$-metric, in order to aid in the construction of
error-correcting codes for the $\ell_\infty$-metric.  By doing so, we
were able to construct codes that achieve better asymptotic rates than
previously known constructions, while also incorporating the property
of being Gray codes. As with previously known constructions, we have
shown that these codes allow for linear-time encoding and decoding of
noisy data.

However, there remains a gap between the best known upper-bound 
for code sizes (either in the general case or in the specific 
case of Gray codes), based on the code-anticode approach presented 
in \cite{TamSch10}, and achievable sizes (both known constructions 
and proven lower-bounds). We therefore propose that more research 
into upper and lower bounds on achievable code sizes is warranted.

Furthermore, much as in the case of codes designed for the Kendall
$\tau$-metric, our auxiliary construction has some asymmetry
between the cases of even- and odd-sized congruence classes. Although 
mostly alleviated by \autoref{aux:final}--in particular for large 
$k$--this creates an irregularity in the slope of the graph of 
asymptotic rate; for rankable codes, certain regions of $\delta$ even 
admit a positive slope, whereby a code with a higher normalized distance 
also has a higher rate.
We posit that, as Holroyd conjectured in \cite{Hol16} for $\cK$-snakes, 
$\Gaux(2n,M)$ codes exist satisfying $M>(2n)!/2-O(n^2)$. This
irregularity is especially pronounced when $2n=6$, where we have 
constructed an auxiliary code of size $178 << 360 = \frac{6!}{2}$. We 
may note, however, that in the case of $2n=4$, the constructed auxiliary 
code of size $8$ can be confirmed to be optimal by a manual search.

Finally, we have presented an adaptation of the solutions discussed 
above to the problem of $(2n,M,\cK)$-snakes, which although not yet 
validating Holroyd's conjecture above, is asymptotically tight.

%--------------- A few commands reference

%%%%%%%%%%%%%%%%%%%%%%%%%%%%%%%%%%%%%%%%%%%%%%%%%%%%%%%
\bibliographystyle{IEEEtranS}
\bibliography{allbib}
%\end{spacing}
%%%%%%%%%%%%%%%%%%%%%%%%%%%%%%%%%%%%%%%%%%%%%%%%%%%%%%%
%%%%%%%%%%%%%%%%%%%%%%%%%%%%%%%%%%%%%%%%%%%%%%%%%%%%%%%%%%%%%%%%%%%%%%%%%%%%%%%
%%%%%%%%%%%%%%%%%%%%%%%%%%%%%%%%%%%%%%%%%%%%%%%%%%%%%%%%%%%%%%%%%%%%%%%%%%%%%%%

%\newpage
%\includepdf[pages={1}]{frontmatter-heb.pdf}

%%%%%%%%%%%%%%%%%%%%%%%%%%%%%%%%%%%%%%%%%%%%%%%%%%%%%%%%%%%%%%%%%%%%%%%%%%%%%%%
%%%%%%%%%%%%%%%%%%%%%%%%%%%%%%%%%%%%%%%%%%%%%%%%%%%%%%%%%%%%%%%%%%%%%%%%%%%%%%%

\end{document}